\documentclass{article}

\usepackage{PRIMEarxiv}

\usepackage[utf8]{inputenc} % allow utf-8 input
\usepackage[T1]{fontenc}    % use 8-bit T1 fonts
\usepackage{hyperref}       % hyperlinks
\usepackage{url}            % simple URL typesetting
\usepackage{booktabs}       % professional-quality tables
\usepackage{caption}
\usepackage{amsfonts}       % blackboard math symbols
\usepackage{nicefrac}       % compact symbols for 1/2, etc.
\usepackage{microtype}      % microtypography
\usepackage{lipsum}
\usepackage{fancyhdr}       % header
\usepackage{graphicx}       % graphics
\usepackage{amsthm,amsmath}
\graphicspath{{media/}}     % organize your images and other figures under media/ folder
\usepackage{natbib} 
\usepackage{threeparttable}
\usepackage{verbatim}
\usepackage{xcolor}
\usepackage{float}
\usepackage{booktabs}
\usepackage{algorithm}
\usepackage{algpseudocode}
\usepackage{multirow}

\usepackage{makecell}        % for table cells
\usepackage[table]{xcolor}
\usepackage{graphicx}
\definecolor{bestyellow}{HTML}{FFFF00}  % table cell colors
\definecolor{secondgreen}{HTML}{D9EAD3} % table cell colors

\theoremstyle{definition}
\newtheorem{example}{Example}[section]
\theoremstyle{remark}

\title{The dangers of using three-number summaries to estimate unknown standard deviations: sensitivity analyses and some possible improvements incorporating shape
}

\author{
  Udara Kumaranathunga,\; Alysha De Livera,\; Luke A. Prendergast* \\
  Department of Mathematical and Physical Sciences \\
  La Trobe University\\
  Melbourne, Australia, 3086\\
  \texttt{*luke.prendergast@latrobe.edu.au} 
}

\begin{document}
\maketitle

\begin{abstract}
In recent years, there has been much progress toward the development of methods for converting three- and five-number summary statistics (i.e. minimum, maximum, median, and quartiles) to means and standard deviations (SDs).  This is commonly done in the meta-analysis setting, where some studies report means and SDs, while other report quantile summaries.  However, we show that three-number summaries, which are the most common, do not contain enough information to reliably estimate SDs.  We show that very poor estimates can result, which may invalidate any inference and provide details of a sensitivity analysis that can allow researchers to have greater confidence in their results, or highlight potential sources of bias.  We further explore whether nominating additional information can provide enough information regarding the unknown data shape to improve SD estimations, and in doing so introduce a new estimator using the scaled Beta distribution.   Simulations and a real data example are used to highlight the advantages and disadvantages of this approach. A Web application is also provided to help researchers perform sensitivity analyses.
\end{abstract}

% keywords can be removed

\section{Introduction}

Over the last few decades, there has been considerable interest in the estimation of sample means and standard deviations (SDs) when only quantile-summary information is available (e.g., median and interquartile range, IQR).  An example where such methods are commonly used is in meta-analysis, where some studies to be synthesized report continuous data outcomes as means and SDs, whereas others report medians and quartiles or sample extremes.  Popular conversion methods include those suggested by \cite{hozo2005estimating}, \cite{wan2014}, \cite{bland2015estimating}, \cite{luo2018}, \cite{shi2020}, and \cite{mcgrath2020}, with new methods being introduced almost every year \citep[e.g.][]{de2024novel, mrakhan2025estimating, tang2026minimum}.  As an indicator of how widely used these methods are, the Wan methods alone have attracted 1,140 citations in Google Scholar in the first 5 months of 2026 (date searched: 1st June, 2026) and more than 11,000 citations in total.

When both the minimum and maximum and the quartiles are available (the so-called \textit{five-number summary}), there exist several methods that can perform well, including the methods described by Wan et al., Shi et al. and McGrath et al. However, when only \textit{three-number summaries} (median and IQR or range intervals) are available, overly restrictive assumptions are unavoidable, and the performance of the methods vary greatly depending on the underlying shape of the unknown data distribution.  With only these three-number summaries, distributional assumptions cannot be adequately tested and the magnitude of estimation error is unknown.

The purpose of this paper is to highlight how sensitive SD estimates are to the methods chosen when only three-number summaries are available.  In Section 2 we review some popular existing methods that will be used later.  In Section 3 we provide numerous examples that highlight the sensitivity of the SD estimator and discuss the dangers of a large error for any analysis that follows.  Motivated by the recent quantile estimation method of \cite{mcgrath2020}, in Section 4 we introduce a goodness-of-fit type sensitivity analysis that researchers can use to assess and report ranges of plausible SD estimates arising from different distribution shapes.  Section 5 concerns the inclusion of additional information, specifically population minima and maxima, to improve estimation.  A new estimator based on the scaled beta (a beta distribution that is not restricted in the domain to $[0,\ 1]$) is proposed.  Performance of the new estimator is considered by simulation in Section 6 where we additionally assess the ability of existing five-number summary methods to estimate the SD when the population extremes replace unknown sample minimum and maximums.  We finish with a summary and recommendations in Section 7 and provide details of a new web application for researchers to perform a sensitivity analysis in Section 8.

\section{Methods}\label{sec:methods}

In this section, we provide details for several methods for estimating mean and SDs from summary data, specifically those that use three-number summaries.

\subsection{Three- and five-number summaries}

When skew and/or outliers are present, some studies may choose to summarize data using sample quantiles instead of the sample mean and SD.  Consider the following three- and five-number summary sets of statistics, with the addition of sample size, $n$,
\begin{align*}
    S_1=&\{a,\ m,\ b,\ n\},\\
    S_2=&\{q_1,\ m,\ q_3,\ n\},\\
    S_3=&\{a,\ q_1,\ m,\ q_3,\ b,\ n\}\\
\end{align*}
where $a,q_1,m,q_3,b$ are the sample minimum, first quartile, median, third quartile, and maximum, respectively, and $n$ is the sample size.   Variants of the above scenarios are also possible.  For example, \cite{mrakhan2025estimating} considered the scenarios in which the sample mean is also reported. Our focus is on estimating the SD for scenarios $S_1$ and $S_2$ and therefore we detail some existing methods commonly used for these scenarios in the following.

\subsection{Assuming underlying normality}

Earlier work by \cite{hozo2005estimating} (scenario $S_1$) and \cite{wan2014} (scenarios $S_1$, $S_2$ and $S_3)$ remain popular choices for methods to estimate means and SDs from the summaries.  As examples, for the estimation of the sample mean, Hozo et al. proposed $\tilde{x}_1=(a + 2m + b)/4$ for $n\leq 25$ and $\tilde{x}_1=m$ for larger samples which assumes underlying symmetry. Later, again under the assumption of symmetry, Wan et al. used $\tilde{x}_1=(q_1 + m + q_3)/3$.  In recent years, in noting the suboptimal use of the sample size and in noting that the weights used for the combining of the summary quantiles may be improved, several researchers have proposed refinements.  It should be noted that although underlying normality is assumed, Wan et al.'s estimators do incorporate sample size and provide good performance if normality is justified.  \cite{luo2018} derived optimal weights for the quantiles for mean estimation and moving forward we favor these while still focusing on the Wan et al. estimators of the SD.    The Luo et al. estimators of means and the Wan et al. estimators of SDs for scenarios $S_1$ and $S_2$ are
\begin{description}
    \item[$S_1$:] $\displaystyle \tilde{x}_1=\left(\frac{4}{4+ n^{3/4}}\right)\frac{(a+b)}{2}+\left(\frac{n^{3/4}}{4+ n^{3/4}}\right)m,\;\;\tilde{s}_1=\frac{b-a}{2\Phi^{-1}\left(\frac{n-3/8}{n+1/4}\right)}$.
    \item[$S_2$:] $\displaystyle \tilde{x}_2=\left(0.7 + \frac{0.39}{n}\right)\frac{(q_1+q_3)}{2}+\left(0.3 - \frac{0.39}{n}\right)m,\;\;\tilde{s}_2=\frac{q_3-q_1}{2\Phi^{-1}\left(\frac{3n/4-1/8}{n+1/4}\right)}$.
\end{description}

For comparative purposes, it is useful to consider the approximate forms of $\tilde{x}_js$ and $\tilde{s}_js$ for large $n$, which can help to understand the dangers of using the methods when the summaries are for data not sampled from normal distributions.  In particular
$$\text{For}\ n\rightarrow \infty:\ \tilde{x}_1 = m,\ \tilde{x}_2=0.7 \frac{(q_1+q_3)}{2}+0.3m.$$

Firstly, for very large $n$, $\tilde{x}_1$ will be approximately equal to the median.  The sample median is known to be a good estimator of the mean of a normal distribution for large sample sizes, and so it makes sense to have less emphasis on the extremes, which can be highly volatile for large sample sizes.  This may indicate the risk of using this approximation if the underlying distribution is very heavy-tailed, in which case $(a+b)$ can be very large and not significantly reduced by using the weight $4/(4 + n^{3/4})$.

Secondly, $\tilde{x}_2$ is approximately a weighted average of the median and average of the quartiles (both of which are estimates of the mean for symmetric distributions).  There is less risk of this approach for heavy-tailed distributions, as the quartiles will be much less volatile than the extremes.  However, it is the available summaries that dictate the choice between these two estimators and not user choice.

Although approximate means are weighted averages of the median, $(q_1+q_3)/2$ or $(a+b)/2$, which are all estimates of the mean for symmetric distributions, SD approximations are very different.  The estimators use a re-scaling of the difference between either $a$ and $b$ or $q_1$ and $q_3$ with the re-scaling factor calculated assuming an underlying normal distribution.  As a consequence, even if symmetry does hold, poor estimates of the SD can result if normality does not.    For later, note that $\lim_{n\rightarrow  \infty} (3n/4 - 1/8)/(n + 1/4)= 3/4$ so that, under some mild conditions,
\begin{equation}
    \tilde{s}_2 \xrightarrow{\text{a.s.}} \frac{Q_3-Q_1}{2\Phi^{-1}(3/4)}\label{eq:a.s.s_2}
\end{equation}
where $Q_1$, $Q_3$ are the population quartiles.  We shall explore this in more detail later.

\subsection{Choosing an underlying distribution using the Quantile Estimation method}\label{section:QE}

Unlike the closed-form estimators above, the quantile estimation (QE) method \citep{mcgrath2020} estimates parameters by solving optimization problems rather than using direct analytic expressions.  The QE method assumes that the outcome follows one of four candidate parametric distributions (normal, lognormal, gamma, Weibull, beta). The parameters are estimated by matching the reported quantiles with the corresponding theoretical quantiles of each candidate distribution. Let $F_{j,\boldsymbol{\theta}_j}$ denote the distribution function of the $j$th candidate distribution with parameter vector $\boldsymbol{\theta}_j$, the QE method estimates $\boldsymbol{\theta}_j$ for the two three-number summary scenarios as
\begin{description}
    \item[$S_1$:] 
    $\displaystyle \hat{\boldsymbol{\theta}}_{1j}:=
    \arg\min_{\boldsymbol{\theta}_j}
    \left[
    \left(F^{-1}_{j,\boldsymbol{\theta}_j}(1/n)-a\right)^2 +
    \left(F^{-1}_{j,\boldsymbol{\theta}_j}(1/2)-m\right)^2+
    \left(F^{-1}_{j,\boldsymbol{\theta}_j}(1-1/n)-b\right)^2
    \right],$     
    \item[$S_2$:] 
    $\displaystyle \hat{\boldsymbol{\theta}}_{2j}:=
    \arg\min_{\boldsymbol{\theta}_j}
    \left[
    \left(F^{-1}_{j,\boldsymbol{\theta}_j}(1/4)-q_1\right)^2+
    \left(F^{-1}_{j,\boldsymbol{\theta}_j}(1/2)-m\right)^2+
    \left(F^{-1}_{j,\boldsymbol{\theta}_j}(3/4)-q_3\right)^2
    \right].$
\end{description}

The best-fitting distribution is the candidate whose parameters estimates minimize the above objective function, and the mean and SD are calculated using properties of the distribution. E.g., if the Gamma distribution is selected and $\hat{\boldsymbol{\theta}}_{2j}=[\widehat{\alpha},\ \widehat{\theta}]^\top$ where $\alpha$ and $\theta$ are the shape and scale parameters, then $\tilde{x}_2=\widehat{\alpha}\cdot\widehat{\theta}$ and $\tilde{s}_2=\sqrt{\widehat{\alpha}\cdot\widehat{\theta}^2}$.  This approach relaxes strict normality or symmetric distribution assumptions, but still relies on parametric assumptions, and its performance depends on whether the true underlying distribution is well approximated by one of the candidates.

\subsection{Using the Box-Cox transformation to achieve symmetry}\label{sect:BC}

\cite{mcgrath2020} also introduced the Box-Cox (BC) method, which assumes that there exists a power parameter $\lambda$ such that the transformed variable $t(X; \lambda)$ of the skewed outcome $X$ is approximately normally distributed.
For $\lambda \in \mathrm{R}$, the Box-Cox transformation is defined as
\[
t(x;\lambda) = 
\begin{cases}
\displaystyle \frac{x^\lambda - 1}{\lambda}, & \text{if } \lambda \ne 0 \\
\log x, & \text{if } \lambda = 0
\end{cases}
\qquad x>0,
\].

Let $Y=t(X;\lambda)$, the BC method assumes that $Y \sim \mathrm{Normal}(\mu_0,\sigma_0^2)$, so that the transformed quantiles are approximately symmetric around the transformed median. That is, for the three-number scenarios, $\lambda$ is selected such that $\displaystyle t(a;\lambda)-t(m;\lambda)\approx t(m;\lambda)-t(b;\lambda)$ for $S_1$ and $\displaystyle t(q_1;\lambda)-t(m;\lambda)\approx t(m;\lambda)-t(q_3;\lambda)$ for $S_2$.

After estimating $\lambda$, the transformed quantiles are treated as approximately normal, and normal-based estimators are applied on the transformed scale to estimate $\mu_0$ and $\sigma_0$. The estimated mean and SD on the original scale are then obtained by applying the inverse Box-Cox transformation and \cite{mcgrath2020} compute these numerically using Monte Carlo simulation. Specifically, random samples are generated from the fitted transformed normal distribution, inverse Box-Cox transformed back to the original scale, and the sample mean and SD of the simulated values are used as estimates of the population mean and SD.

When it is reasonable to assume at least approximate normality following the BC transformation, e.g. log-normal data when a log transformation leads to the normal distribution, the method can substantially reduce the estimation error relative to directly applying normal-based formulas to the raw quantiles. However, if the underlying distribution is highly skewed or heavy-tailed, there might not be a $\lambda$ to achieve approximate normality. As we shall soon see in the next section, SD estimation ultimately depends on the shape of the distribution (spacing between quantiles and extremes), under- or over-estimation of tail spread on the transformed scale can remain after back-transformation. Moreover, as the inverse transformation is nonlinear and small discrepancies in tail behavior under the transformed scale may produce large differences in the estimated SD after inversion.

\section{Implications of SD estimation using three-number summaries}
\label{sec:3}

Even if the underlying distribution is symmetric, there is not enough information in a three-number summary to draw any conclusions about the shape of the distribution. Additionally, $(Q_3-M)=(M-Q_1)$ (and similarly for min, max, and median), does not guarantee symmetry more broadly for an entire distribution.  It therefore suffices at this point to study simple shapes of symmetric distributions to understand the implications of assuming normality, while noting that there are other cases (e.g. non-symmetric) for which results may be even worse.  Hence, our first study of symmetric distributions provides insight into both the approximated SD from the raw quantiles and also following the use of transformations for symmetrization (e.g. the BC).  We then provide simulation studies to assess the performance of existing methods for various skewed and non-skewed distributions.  We finish this section with a consideration on symmetrization before studying the impact of poor SD approximations on statistical inference.

\subsection{The influence of shape on SD approximations for symmetric distributions}

We begin with a simple example.   Simplicity should not downplay the relevance of this example, since uniform distributions may not only approximate some measurements, but uniformity can also follow symmetrization, as we see later with the Pareto distribution.

\begin{example}
    Let $X\sim \text{Uniform}(0,1)$ so that $Q_1=1/4$, $M=1/3$ $Q_3=3/4$ and the mean and SD are $\mu=E(X)=1/2$ and $\sigma =\sqrt{\text{Var}(X)}=1/\sqrt{12}$ ($\approx 0.289$).  If we erroneously assume a underlying normal distribution, our true approximate SD, assuming large $n$ (specifically, $n\rightarrow \infty$) and and using \eqref{eq:a.s.s_2} is $$\tilde{\sigma}=\frac{Q_3-Q_1}{2\Phi^{-1}(3/4)}\approx 0.3706.$$  The relative error (RE) expressed as a percentage is almost 29\%.  As expected due to symmetry, the approximated mean $\tilde{\mu}=1/2$ is exactly equal to the target mean (i.e., RE$=0$).
\end{example}

\begin{example}

We now extend our study of the uniform distribution to other symmetric distributions.
    
\begin{table}[ht]
  \centering
  \begin{minipage}{0.6\linewidth}
    \centering
    \caption{Large $n$ approximated SDs ($\tilde{\sigma}$) and relative error (RE) in the approximation when compared to the true SD ($\sigma$).}\label{tab:REs}
    \begin{tabular}{lllll}
      \toprule
      Distribution  & $\sigma$ & $\widetilde{\sigma}$ & RE & Comments\\
      \midrule
      $N(\mu, \sigma^2)$ & $\sigma$  & $\sigma$ &  0\% & Ideal scenario\\
      $t_3$  & $1.732$  & $1.134$ &  -34.53\% & Heavy-tailed \\
      $t_6$  & $1.225$  & $1.064$ &  -13.14\% & Less heavy-tailed \\
      $t_{12}$  & $1.096$  & $1.031$ &  -5.87\% & Approaching normal \\
      Uniform(0,1)  & $0.289$  & $0.371$ &  28.4\% & Uniform \\
      Uniform(0,10)  & $2.887$  & $3.706$ &  28.4\% & Uniform \\
      Beta$(0.5,0.5)$  & $0.354$  & $0.524$ &  48.3\% & $U$-shaped \\
      Logistic$(0,1)$  & $0.901$  & $0.814$ &  -10.2\% & Peaked \\
      \bottomrule
    \end{tabular}
  \end{minipage}
\end{table}

Relative errors for SD approximations from various symmetric distributions, assuming an underlying normal distribution, are given in Table \ref{tab:REs}.  Obviously, the ideal scenario is that the true distribution is normal, in which case there is zero error.  However, despite the symmetry of the studied distributions, approximation error can be large, ranging from nearly -35\% for the heavy-tailed $t_3$ distribution to almost 50\% for the U-shaped Beta$(0.5,0.5)$.  Therefore, even if symmetrization is achieved, large approximation errors can result when the shape of the symmetrized distribution is non-normal. E.g., and as we shall discuss more later, symmetrizing a Pareto distribution results in uniformity, and the expected RE is more than 28\%.

\end{example}

When we have just symmetric quartiles (or median and extremes), there is no way of knowing the underlying shape.  In the next example, we consider the differences in approximate SDs for different distributions with the same quartiles.

\begin{example}

We now consider a hypothetical example where we have quartiles $Q_1=1/4$, $Q_2=1/2$ and $Q_3=3/4$. There are many candidate distributions that would fit this quartile profile, the most obvious being Uniform(0, 1).

\begin{table}[ht]
    \centering
    \begin{tabular}{llllll}
    \toprule
     Distribution & SB$(L, U, 1/2, 1/2)$ & Uniform($a$, $b$) & N$(\mu, \sigma^2)$ & Logistic$(\mu, \sigma^2)$ & GT$(\mu, \sigma, df = 3)$ \\
     Parameters & 0.146, \ 0.853  & 0, \ 1 & 1/2,\ 0.370 & 1/2, \ $[4\text{ln}(3)]^{-1}$ & 1/2, \ 0.327\\
     SD  & 0.250 & 0.289 & 0.371 & 0.413 & 0.566\\
     Shape & $U$-shaped & Uniform & Gaussian & Peaked & Heavy-tailed \\
     \bottomrule
    \end{tabular}
    \caption{Examples of distributions that provide $Q_1=1/4$, $Q_2=1/2$ and $Q_3=3/4$, the parmeters needed to achieve these quartiles, the resulting SD for the distribution and a brief note on the shape.  Considered are the scaled Beta (SB), uniform, normal, logistic and a generalised $t$ (location shifted and a scale factor).}
    \label{tab:diff_dist_diff_SD}
\end{table}

In Table \ref{tab:diff_dist_diff_SD} we provide  examples of distributions that have this quartile profile. In addition to the uniform, normal, and logistic distributions we considered earlier, we also include the scaled (four parameter) beta (SB) distribution, which is defined on the domain $[L,\ U]$ using a location shift and scaling, and a generalized $t$ (GT) distribution, which also includes a location shift and a scale factor.  The SDs for these distributions can be very different, ranging from 0.25 to 0.566.  This again highlights that the three number summary is not enough on its own to inform an SD. E.g., if an underlying normal was assumed, then this could be either an over- or under-approximation of the target SD.

\end{example}

Our examples so far highlight that even when symmetry is evident, the SD approximation can be severely compromised when a normal distribution is assumed erroneously.   While we studied case $S_2$ (quartiles and sample size), scenario $S_1$ (median and extremes) also suffers from these same issues.

\subsection{Simulations assessing SD estimation using existing methods}

We now turn our attention to the performance of existing methods in approximating the SD from the three-number summary and further include some non-symmetric distributions, as well as consider both the $S_1$ and $S_2$ scenarios.

\begin{example}

For this example, we consider the performance of several existing methods when data is sampled from a normal distribution (the ideal) and then other symmetric and non-symmetric distributions.  For varying sample sizes, $n$, we sampled 1000 data sets independently, estimated summary measures minimum, first quartile, median, third quartile, and the maximum to obtain the three-number summaries $S_1$ and $S_2$ for each data set.  We then applied the Want, QE and BC methods to each of these scenarios to estimated (approximate) the sample SD.  We compare the resulting SD to the sample SD (the target) and reported the average relative error across the simulated data sets.

\begin{table}[ht]
\centering
%\small
\begin{tabular}{c l r r r r r r}
\toprule
$n$ & Distribution & Wan$_{S1}$ & QE$_{S1}$ & BC$_{S1}$ & Wan$_{S2}$ & QE$_{S2}$ & BC$_{S2}$ \\
\midrule
50 & ${N}(50,17^2)$ & 1.11\% & 10.41\% & -3.30\% & 0.89\% & -0.19\% & -7.04\% \\
50 & $t_{3}$ & 28.62\% & 41.46\% & 17.16\% & -25.75\% & -19.86\% & -45.59\% \\
50 & $t_{6}$ & 13.77\% & 25.05\% & 6.70\% & -11.12\% & -5.39\% & -33.89\% \\
50 & ${U}(0,10)$ & -25.43\% & -18.69\% & -30.85\% & 26.54\% & 32.83\% & 0.42\% \\
50 & $\text{Beta}(0.5,0.5)$ & -36.90\% & -1.27\% & -43.06\% & 44.45\% & 59.60\% & -1.24\% \\
50 & $\text{Beta}(0.8,0.2)$ & -24.19\% & 4.74\% & -66.74\% & -18.30\% & -3.94\% & -38.67\% \\
50 & $\text{Logistic}(0,1)$ & 10.05\% & 20.22\% & 2.50\% & -8.44\% & 0.19\% & -38.25\% \\
50 & $\text{LogN}(4,0.3)$ & 2.53\% & 13.91\% & -1.50\% & -4.74\% & -4.63\% & -5.32\% \\
50 & $\text{SN}(0,1,-10)$ & -5.32\% & 1.10\% & -28.34\% & 2.60\% & 1.48\% & -35.87\% \\
50 & $\text{SN}(0,1,10)$ & -5.73\% & 5.28\% & -8.24\% & 3.67\% & 29.51\% & 7.08\% \\
50 & $\text{Half-N}(0,1)$ & -7.92\% & 4.02\% & -8.68\% & 3.08\% & 27.84\% & 7.43\% \\
50 & $\text{Pareto}(10,3)$ & 15.40\% & 44.04\% & -25.87\% & -42.00\% & -39.68\% & -39.80\% \\
\midrule
200 & ${N}(50,17^2)$ & 0.78\% & 6.81\% & -1.66\% & 0.15\% & -0.09\% & -4.23\% \\
200 & $t_{3}$ & 67.27\% & 76.14\% & 56.30\% & -30.06\% & -29.45\% & -56.61\% \\
200 & $t_{6}$ & 28.35\% & 37.04\% & 23.63\% & -12.47\% & -11.57\% & -44.08\% \\
200 & ${U}(0,10)$ & -37.13\% & -33.26\% & -38.90\% & 27.97\% & 28.75\% & -7.93\% \\
200 & $\text{Beta}(0.5,0.5)$ & -48.27\% & -10.12\% & -49.83\% & 47.57\% & 51.32\% & -18.03\% \\
200 & $\text{Beta}(0.8,0.2)$ & -35.73\% & 1.79\% & -74.83\% & -18.01\% & -1.34\% & -36.13\% \\
200 & $\text{Logistic}(0,1)$ & 19.04\% & 25.90\% & 15.28\% & -9.57\% & -8.66\% & -50.22\% \\
200 & $\text{LogN}(4,0.3)$ & 6.15\% & 10.29\% & -0.69\% & -5.79\% & -2.73\% & -2.82\% \\
200 & $\text{SN}(0,1,-10)$ & -6.48\% & -1.88\% & -32.96\% & 1.82\% & 1.12\% & -39.93\% \\
200 & $\text{SN}(0,1,10)$ & -5.90\% & -4.06\% & -14.60\% & 1.48\% & 16.44\% & -0.47\% \\
200 & $\text{Half-N}(0,1)$ & -10.58\% & -3.49\% & -11.70\% & 2.71\% & 17.67\% & 0.45\% \\
200 & $\text{Pareto}(10,3)$ & 52.95\% & 60.65\% & -30.67\% & -50.78\% & -47.65\% & -48.98\% \\
\bottomrule
\end{tabular}
\caption{Percentage relative errors (RE) of estimating SD using existing methods under $S_1$ and $S_2$ for different distributions and sample sizes.  Note that LogN, SN, Half-N denote the lognormal, skew normal and half normal distributions respectively.}\label{tab:sim_existing_methods}
\end{table}

We provide the empirical REs in Table \ref{tab:sim_existing_methods} for two choices of sample size, $n=50$ and $n=200$.  When data is sampled from the normal distribution (ideal), the REs are typically small, with Wan performing best for both the $S_1$ and $S_2$ scenarios.  QE suffered somewhat with scenario $S_1$, suggesting that it is difficult for some simulated data sets to choose the correct distribution using minimum and maximum.  However, the QE approach performed very well for $S_2$.  With symmetrization not necessary for data sampled from the normal, BC was outperformed by the Wan method.

For other distributions, both symmetric and non-symmetric, performance varied greatly and in many cases performance deteriorated with increasing sample size.  Each of the methods has reasonable performance for at least some of the distributions; however, there was no non-normal distribution for which all methods performed well.  The very large differences in REs across the methods and distributions studied highlight that with only a three-number summary, it would be difficult to have confidence in estimated SDs in practice.

\end{example}

\subsection{The effect of symmetrization}

Using the Box-Cox transformation to symmetrize the quantiles \citep{mcgrath2020} is a promising approach, as it may help to diminish the impact of extreme skew; see Section \ref{sect:BC} for details.   Even if symmetry is achieved, as we have shown previously in this section, very poor estimated SDs can still result when assuming normality.  We previously showed in Table \ref{tab:sim_existing_methods} that existing methods for approximating SD provide poor approximations for the Pareto and Skew-Normal distributions, among others.   In Figure \ref{fig:par_and_sk_dist} of the Supplementary Information, we provide probability density functions for the Pareto distribution with shape 2.1, 3 and 5 (Plots 1 to 3) and the Skew-Normal with 5, 10 and 20 (Plots 4 to 6).  The Pareto distribution has finite variance for $\alpha > 2$ which is why we chose values greater than 2 only.

\begin{figure}[ht]
    \centering
    \includegraphics[width=0.75\linewidth]{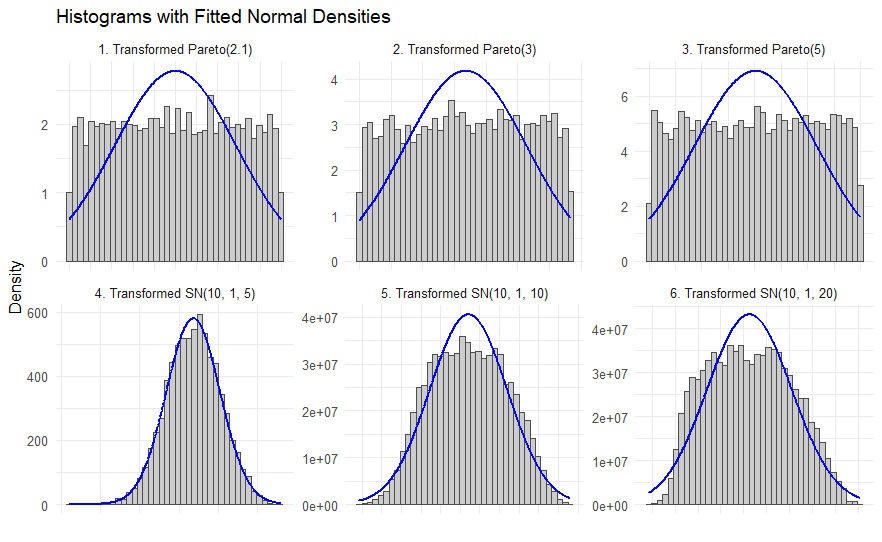}
    \caption{Histograms of 10,000 Box-Cox transformed values from various Pareto and Skew-Normal distributions where the transformation parameter was chosen so that the transformed population first quartile, median and 3rd quartile were as close to being symmetric as possible.  The blue line is the fitted normal density based on the sample mean and SD of the 10,000 data points.}
    \label{fig:transformed_hist}
\end{figure}

We used the smoothed Box-Cox functionality from the `estmeansd' package \citep{mcgrath2019estmeansd} to find the optimal transformation parameter.  While the function searchers over $[-2,\ 3]$, we used $[-10,\ 10]$ since the optimal value was outside the default range for some of these distributions.  In Figure \ref{fig:transformed_hist} we provide histograms of 10,000 data points generated from the symmetrized distributions.  For the Pareto distribution, the histograms are approximately uniform.  This is expected since the quantile function for the Pareto distribution is $Q(p)=(1-p)^{-1/\alpha}$ and the transformation parameter equal to $-\alpha$ results in $\left\{\left[Q(p)\right]^{-\alpha} - 1\right\}/(-\alpha)=p/\alpha$ which is the quantile function of a uniform distribution.  Hence, while symmetry is achieved, the result is not close to normal.  For the Skew-Normal distributions, at least approximate symmetry is also achieved, and for the smaller slant parameter value of 5 the result is approximately normal.  However, for $\alpha=10,20$, while approximate symmetry is evident, the shape is non-normal. 

\subsection{Implications of poor SD estimation for statistical inference}

Naturally, a poor estimate of the sample SD can result in invalid statistical inference. Although there are many ways in which an approximated SD may be used for testing, we choose the commonly used mean difference and associated two-sample $t$-test and Cohen's $d$ test to highlight the impact of poor approximations.  Let $\overline{x}_i,s_i,n_i$ $(i=1,2)$ denote the sample mean, standard deviation, and sample size for two independent samples.  The mean difference and its standard error (SE) are
\begin{equation}
    \overline{x}_1-\overline{x}_2,\ \text{SE}=\sqrt{\frac{s_1^2}{n_1} + \frac{s_2^2}{n_2}}
\end{equation}
and, for  Cohen's $d$ \citep[see, e.g.][]{Borenstein2009book},
\begin{equation}
    d = \frac{\overline{x}_1-\overline{x}_2}{s_p},\\ \text{SE}=\frac{N}{n_1n_2} + \frac{d^2}{2N}
\end{equation}
where $N=n_1+n_2$ and $s_p^2=[(n_1-1)s_1^2+(n_2-1)s_2^2]/(n_1+n_2-2)$ is the pooled sample variance.

\begin{example}
    Consider a hypothetical example where $\overline{x}_1=30$, $\overline{x}_2=50$, $s_1=s_2=15$ $n_1=n_2=30$ so that for Cohen's $d$ a small to moderate effect of $d=1/3$ (\text{SE}=0.26) is detected.  Suppose now that the sample means are not changed (e.g. have been well approximated, which would be expected when there is underlying symmetry) but that the sample $s_i$s are under-approximated by a factor of $c$ (which can occur when there is underlying symmetry but non-normal).  That is, we consider the impact of \
    $$s_i^* = c\times s_i,\ (i=1,2)$$
    where, e.g. $c=1$ indicates zero underestimation and $c=0.7$ gives 30\% under-approximation.

\begin{table}[ht]
\centering
\begin{tabular}{c c c c l c c c l}
\toprule
 & \multicolumn{4}{c}{Mean difference} & \multicolumn{4}{c}{Cohen's $d$} \\
\cmidrule(lr){2-5} \cmidrule(lr){6-9}
$c$ & $\overline{x}_1-\overline{x}_2$ & SE & 95\% CI & p-value & $d$ & SE & CI & p-value \\
\midrule
1.0 & 5 & 3.87 & (-2.75, 12.75) & 0.202 & 0.33 & 0.26 & (-0.18, 0.84) & 0.200 \\
0.9 & 5 & 3.49 & (-1.98, 11.98) & 0.157 & 0.37 & 0.26 & (-0.14, 0.88) & 0.155 \\
0.8 & 5 & 3.10 & (-1.20, 11.20) & 0.112 & 0.42 & 0.26 & (-0.09, 0.93) & 0.110 \\
0.7 & 5 & 2.71 & (-0.43, 10.43) & 0.070 & 0.48 & 0.26 & (-0.04, 0.99) & 0.069 \\
0.6 & 5 & 2.32 & (0.35, 9.65)* & 0.036* & 0.56 & 0.26 & (0.04, 1.07)* & 0.035* \\
0.5 & 5 & 1.94 & (1.12, 8.88)** & 0.012** & 0.67 & 0.26 & (0.15, 1.19)** & 0.012** \\
\bottomrule
\end{tabular}
\caption{Results for Cohen's $d$ and mean differences including standard errors (SEs), 95\% confidence intervals and p-values when the sample $s_i$s are multiplied by $c$.  * and ** denote strength of siginiface against a null hypothesis of zero at the 5\% level.}
\label{tab:md_and_d_results}
\end{table}

In Table \ref{tab:md_and_d_results} we provide the results for $c=1,\dots, 0.5$.  Since $c=1$ gives zero under-approximation of the sample SDs, this row is the reference row for comparison with the other choices of $c$.  When $c=1$, the effect detected using Cohen's $d$ is small to moderate, and for both the two-sample $t$-test and the test for $d$, comparisons against a null of zero are not significant at the level of 5\%. Under-approximating the sample SDs affects the mean difference and Cohen's $d$ in different ways, although it has a similar impact on the statistical inference of the two (since the $n_i$s are not too small so that the strength of evidence against the null is similar). As expected, the estimated mean difference does not change; however, the SE changes by the factor of $c$.  This means that the strength against the null increases as $c$ decreases and that statistical significance can be erroneously achieved.  The 95\% confidence intervals are also very sensitive to the under-approximation and can lead to very different conclusions when compared to the perfect approximation $(c=1)$.  For Cohen's $d$ the SE remains approximately unchanged, since it is approximately $N/(n_1n_2)$, although the effect size increases from small-to-moderate towards large.  This in itself can lead to erroneous conclusions.  As with the two-sample $t$-test, inference is also highly sensitive.
    
\end{example}

The examples in this section highlight several very important points that researchers need to consider when using three-number summaries to estimate a sample SD.  
\begin{itemize}
    \item Even when an underlying distribution is symmetric, the assumption that it is normal is too strict an assumption, and deviations away from normality can result in very poor SD results.
    \item Symmetrization using the transformation may help to improve results in the presence of skew; however, the assumption of normality post-transformation can again lead to poor results when normality is not achieved. E.g., transforming data sampled from a Pareto distribution leads to uniformity.
    \item It is not possible to identify whether normality assumptions are reasonable from a summary of just three numbers alone.
    \item Under- and over-approximation/estimation can be very large, and it is not possible to predict which from a three-number summary (although obvious skew in the quantiles may indicate under-estimation assuming normality).
    \item The impact of a poorly estimated sample SD from three-number summaries on statistical inference can be large.  Poor results could lead to inflated or deflated effects, unreliable confidence interval coverage, and poorly controlled Type II error. 
\end{itemize}

Based on our findings, we now consider some possibilities for improving the results.

\section{Sensitivity analysis when using three-number summaries}

In practice, a sensitivity analysis should be conducted routinely when modeling assumptions cannot be tested or empirically verified. Such analyses may strengthen confidence in the findings or reveal potential biases.  For example, \cite{Design_of_Obs_studies} repeatedly emphasizes the importance of sensitivity analyses to look for potential hidden biases and for testing conclusions against model assumptions.  With just a three-number summary and without any additional information, restrictive assumptions are unavoidable.  As we have seen and through further illustration in this section, even modest changes in assumptions can lead to very different outcomes.  The approach to this sensitivity analysis is motivated by the QE approach of \cite{mcgrath2020} discussed in Section \ref{section:QE}.

\subsection{Details and methods}

Consider a distribution function, $F$, and where the distribution has two unknown parameters, $\theta_1$ and $\theta_2$.  Based on the three-number summary, we consider estimates of $\theta_1$ and $\theta_2$ and denote these as
\begin{equation}
    \widehat{\boldsymbol{\theta}}_F = [\widehat{\theta}_1,\widehat{\theta}_2]^\top= g_F(q_1,m,q_3)
\end{equation}
where $g_F$ is a function of the summary data that computes the estimates.  The QE approach uses numerical optimization to minimize the sum of squared errors (SSE) between the observed and theoretical quantiles to fit the distribution parameters for scenarios $S_1$, $S_2$, or $S_3$.  We similarly do this for several distributions, including those considered by QE, with starting values for the optimization derived from the estimated quantiles and properties of distribution based on theoretical quartiles.  Details of the starting values for the optimization process are provided in Section \ref{section:optim} of the Supplementary Information.

\begin{algorithm}[htbp]
\caption{Bootstrap algorithm for distribution matching to three-number summary scenario $S_2$.}
\label{alg:bootstrap-quartile-matching}
\begin{algorithmic}[1]
\State $(\widehat{\theta}_{1,0},\widehat{\theta}_{2,0}) \gets \theta_F(q_1,m,q_3; \theta_1,\theta_2)$
\State $Q_{1,0} \gets Q_F(1/4; \widehat{\theta}_{1,0},\widehat{\theta}_{2,0})$
\State $M_0 \gets Q_F(1/2; \widehat{\theta}_{1,0},\widehat{\theta}_{2,0})$
\State $Q_{3,0} \gets Q_F(3/4; \widehat{\theta}_{1,0},\widehat{\theta}_{2,0})$
\State $\mathrm{SSE}_0 \gets (Q_{1,0}-q_1)^2 + (M_0-m)^2 + (Q_{3,0}-q_3)^2$

\For{$i = 1,\ldots,B$}
    \State Generate $n$ observations from $F(\widehat{\theta}_{1,0},\widehat{\theta}_{2,0})$
    \State Estimate the sample quartiles $(q_{1,i}, m_i, q_{3,i})$
    \State $(\widehat{\theta}_{1,i},\widehat{\theta}_{2,i}) \gets \theta_F(q_{i,1},m_i,q_{i,3}; \theta_1,\theta_2)$
    \State $Q_{1,i} \gets Q_F(1/4; \widehat{\theta}_{1,i},\widehat{\theta}_{2,i})$
    \State $M_i \gets Q_F(1/2; \widehat{\theta}_{1,i},\widehat{\theta}_{2,i})$
    \State $Q_{i,3} \gets Q_F(3/4; \widehat{\theta}_{1,i},\widehat{\theta}_{2,i})$
    \State $\mathrm{SSE}_i \gets (Q_{1,i}-q_{1,i})^2 + (M_i-m_i)^2 + (Q_{3,i}-q_{3,i})^2$
\EndFor

\State \Return $\left(\mathrm{SSE}_0,\mathrm{SSE}_1,\ldots,\mathrm{SSE}_B\right)$
\end{algorithmic}
\end{algorithm}

To help assess whether a candidate distribution is a potential match for an observed set of quartiles, we carried out a parametric bootstrap analysis based on the SSEs in the form of a goodness-of-fit test.  The procedure is detailed in Algorithm \ref{alg:bootstrap-quartile-matching} for scenario $S_2$ where we repeatedly generate $B$ bootstrap samples using the parameter estimates from the observed sample quartiles.  For each bootstrap sample, we compute the sample quartiles, re-estimate the distributional  parameters using these quartiles and calculate the SSE by comparing the theoretical quartiles under the refitted distribution with the corresponding sample quartiles.  The p-value from this bootstrapping process is then   
\begin{equation}
\hat p_B = \frac{1}{B} \sum_{i=1}^B \mathbf{1}\!\left\{ \mathrm{SSE}_i \ge \mathrm{SSE}_0 \right\}
\end{equation}
where $\mathbf{1}(\cdot)$ denotes the indicator function that is equal to one if its argument is true and zero otherwise.  With three quartiles holding limited information on potential distributions, we are not suggesting to choose a distribution based on the initial SSE or the bootstrap, but to instead use this information to potentially rule out a distribution and focus the sensitivity analysis on those which seem plausible.

All fitted symmetric distributions will have the same SSE since the same symmetric theoretical quartiles that provide the minimum SSE will be found.  However, the bootstrap results will differ and a larger $\hat p_B$  may result if the degree of asymmetry is more likely for one of the distributions due to less variability in the quartiles.  Similarly, an asymmetric distribution with the smallest SSE may not necessarily have the largest $\hat p_B $. Therefore, it makes sense to compare both the SSEs and the $\hat p_B $s that arise from the bootstrapping.

Finally, it is also possible to perform this bootstrap for scenario $S_1$, and we do so by example later.  To do this, we replace $q_1$ and $q_3$ by the sample minimum and maximum, and $1/4$ and $3/4$ by $1/n$ and $1-1/n$ in the theoretical quantile weights of Algorithm \ref{alg:bootstrap-quartile-matching}.

\subsection{Examples}

We start with a hypothetical example to assess the performance of the SD estimators with increasing skewness in the three-number summary, together with the estimated SDs from fitting other distributions and the bootstrapping goodness-of-fit process.  We will follow that with some analyses of real data.

\begin{example}
    
  For this example, we consider the symmetric quartiles $q_1=2$, $m=10$, and $q_3=18$ and introduce asymmetry by adding a constant $b$ to $q_3$ (i.e., $q_3 + b$).

\begin{table}[ht]
\centering
\small
\begin{tabular}{llcccc}
\toprule
\textbf{Method} & $b=0$ & $b=2$ & $b=4$ & $b=8$ & $b=16$ \\
\midrule
Wan     & 12.21 & 13.74 & 15.27 & 18.32 & 24.43 \\
BC      & 5.52 & 6.72 & 8.36 & 13.46 & 47.83 \\
QE      & 11.86 & 13.34 & 19.49 & 25.53 & 59.36 \\
Normal    & 11.86 (0, 1) & 13.34 (0.67, 0.563) & 14.83 (2.67, 0.273) & 17.79 (10.67, 0.054) & 23.72 (42.67, 0.005) \\
Uniform    & 9.24 (0, 1) & 10.39 (0.67, 0.542) & 11.55 (2.67, 0.271) & 13.86 (10.67, 0.058) & 18.47 (42.67, 0.008) \\
Logistic   & 13.21 (0, 1) & 14.86 (0.67, 0.517) & 16.51 (2.67, 0.243) & 19.81 (10.67, 0.055) & 26.42 (42.67, 0.009) \\
LogN   & 28.18 (6.08, 0.003) & 38.91 (4.75, 0.014) & 53.26 (3.69, 0.03) & 98.38 (2.19, 0.138) & 320.67 (0.66, 0.461)\\ Gamma   & 14.14 (3.56, 0.041) & 16.76 (2.27, 0.129) & 19.6 (1.35, 0.256) & 25.92 (0.33, 0.619) & 41.28 (0.08, 0.802) \\
Weibull & 14.25 (3.7, 0.027) & 17.57 (2.54, 0.073) & 21.41 (1.7, 0.161) & 30.94 (0.66, 0.422) & 59.36 (0.01, 0.914) \\
\bottomrule
\end{tabular}
\caption{SD estimates and assessment criteria based on quartiles $q_1=2$, $m=10$ and $q_3=18+b$ for varying $b$ where estimates were obtained using either the Wan, BC or QE methods, or determined using specific distributions.  SSE and bootstrap p-values are in parentheses.}\label{tab:varying_sd}
\end{table}

The SD estimated using the Wan, BC and QE approaches, as well as the SDs assuming a specific distribution with distribution parameters, SSEs, and $\hat p_B $ values estimated using the optimization procedure, are shown in Table \ref{tab:varying_sd}. BC and QE estimates were calculated using the `estmeansd' package \citep{mcgrath2019estmeansd}.  For the symmetric case $(b=0)$, the Wan, BC and QE approaches vary significantly ranging between 5.5 and 12.2.  There is no information in the quantiles that can lead to any suggested preferred method for these symmetric quartiles and, therefore, the results are highly dependent on the choice of the researcher.  When considering specific distributions, the high SSEs and small p-values allow us to rule out asymmetric distributions.  However, there is no power in the quartiles to choose between symmetric distributions, and the estimated SDs vary between 9.24 and 13.2.

As we increase asymmetry, we continue to see variation in the estimated SDs, and in many cases it is difficult to determine which of the SDs may be suitable.  As an example, for $b=4$ giving the quartile set $(2,\ 10, 24)$ we can rule out lognormal due to a high comparative SSE and small $\hat p_B $, although it becomes more difficult to choose between the other options.  The QE approach selects the gamma distribution, although other distributions cannot be discounted and provide notably different SDs.  The BC approach gives the smallest SD of 8.36, and the largest of the other potential choices (not including the lognormal) is close to 18 for the Weibull.

Given that there is nothing extreme in these quartiles that have been considered and also the variation in estimated SDs, this example highlights the need to carry out sensitivity analyses to better understand the implications of the chosen method.
\end{example}

\begin{example}
    We now consider an example of length of stay in hospital but where the sample mean and standard deviation are known, together with the median and IQR.  Knowing the mean and SD will allow us to compare the estimates obtained from the three-number summary with the true sample estimates.  

\begin{table}[ht]
    \centering
    \begin{tabular}{lllllll}
    \toprule
       Group  & $n$ & $\overline{x}$ & $s$ & $q_1$ & $m$ & $q3$  \\ \midrule
       ERAS  & 37 & 20.1 & 5.4 & 15.5 & 19.0 & 25.0 \\
       Control & 37 & 26.8 & 13.5 & 21.0 & 23.0 & 29.5 \\ \bottomrule
    \end{tabular}
    \caption{Sample sizes ($n$), samples means and SDs ($\overline{x}$ and $s$) and sample quartiles ($q_1$, $m$ and $q_3)$ for two groups, ERAS and control, for the \cite{takagi2019effect} study.}
    \label{tab:length_of_stay_data}
\end{table}

\cite{takagi2019effect} considered the impact of a Enhanced Recovery After Surgery (ERAS) protocol after pancreaticoduodenectomy.  The primary outcome was the length of stay (days) in the hospital, and the sample size, means and SD of the samples ($s$) and the sample quartiles for two groups, ERAS and control, are shown in Table \ref{tab:length_of_stay_data}.   For an expanded sensitivity analysis, we include several other distributions that provide different shape; namely the Pareto, inverse Gaussian (inv-Gauss), and log Logistic (LogLogistic).

    \begin{table}[ht]
\centering
\begin{tabular}{lrrrrrrrr}
\toprule
& \multicolumn{4}{c}{ERAS Group} & \multicolumn{4}{c}{Control Group} \\
\cmidrule(lr){2-5} \cmidrule(lr){6-9}
Method & Mean & SD & SSE & $\hat{p}_B$ & Mean & SD & SSE & $\hat{p}_B$ \\
\midrule
Luo/Wan      & 19.89 & 7.33 &  &  & 24.60  & 6.56 &  &  \\
BC       & 21.60  & 8.55 &  &  & 24.86 & 6.53 &  &  \\
QE       & 20.76 & 7.79 &  &  & 25.10  & 6.78 &  &  \\
Normal     & 19.83 & 7.04 & 1.04 & 0.188 & 24.50  & 6.30  & 3.38 & 0.020 \\
Uniform     & 19.83 & 5.49 & 1.04 & 0.204 & 24.50  & 4.91 & 3.38 & 0.016 \\
Logistic    & 19.83 & 7.84 & 1.04 & 0.214 & 24.50  & 7.02 & 3.38 & 0.020 \\
LogN    & 20.76 & 7.79 & 0.30  & 0.480  & 25.10  & 6.78 & 2.33 & 0.044 \\
Gamma    & 20.43 & 7.28 & 0.50  & 0.400   & 24.89 & 6.50  & 2.66 & 0.044 \\
Weibull  & 19.91 & 6.86 & 1.02 & 0.218 & 24.34 & 6.23 & 3.78 & 0.016 \\
Pareto   & - & - & - & - & 27.70  & 14.82 & 0.41 & 0.326 \\
inv-Gauss & 20.75 & 7.67 & 0.29 & 0.518 & 25.10  & 6.73 & 2.32 & 0.070 \\
LogLogistic   & 21.12 & 9.52 & 0.30  & 0.50   & 25.33 & 7.91 & 2.33 & 0.048 \\
\bottomrule
\end{tabular}
\caption{Estimated mean and SDs for length of stay for ERAS and control groups from the \cite{takagi2019effect} study using different methods/distribution assumptions.  The SSEs and $p$-values ($\hat{p}_B$) for the bootstrap goodness-of-fit tests for the different distributions are also shown. }
\label{tab:takagi_estimated_means_and_sds}
\end{table}
\end{example}

We start with the ERAS group.  Recall that the reported sample mean and SD are 20.1 and 5.4.  While the Luo, BC and QE estimate the mean reasonably well, they each overestimate the SD by more than 35\% (and up to 68\%).  The goodness-of-fit test has difficulty identifying suitable distributions. Although the Pareto distribution was not fitted due to the lack of skew in the quartiles, all of the other distributions are potential candidates, with SD estimates varying from 5.49 up to 9.52.  We can see that the reported SD of 5.4 is close to the lower boundary of estimated SDs.

For the control group, the estimated SD is vastly underestimated (approximately 50\%) by almost all methods (Wan, BC, and QE, as well as most distributions).  However, the small bootstrap p-values for all but the Pareto distribution indicate a lack of confidence that these distributions are suitable choices.  All other distributions except for the inverse Gaussian, which reported a p-value of 0.07, were rejected. For Pareto (p-value 0.326), we can see that it is the only choice that reports an estimated SD close to the reported SD (14.82 versus 13.5) and is also the only distribution that estimates the sample mean close to the reported 26.8 (estimated 27.1).  

This example highlights the importance of a sensitivity analysis.  The quantiles reported by \cite{takagi2019effect} were difficult to fit using existing methods and this was highlighted by the sensitivity analysis in the ERAS group and the clear choice of the Pareto distribution for fitting to the control group.

\begin{example}\label{example:real_sensitivity}

We now consider sensitivity analyses for three more real data sets found in the literature.  \cite{yin2021circulating} present a meta-analysis of interleukin-17 (IL-17) levels and disease activity (active vs inactive) in patients with systemic lupus erythematosus (SLE).  Several studies used in the meta-analysis published medians and interquartile ranges (scenario $s_2$), or medians and ranges ($s_1$).  The authors used the Wan method to estimate the unknown means and SDs from these summary measures.  As part of this example, we will consider the summaries used in the meta-analysis published by \cite{lozovoy2014hypertension} and \cite{abo2021cytokines}.  Our third example is based on data from the `metamedian' package \citep{metamedian} which was extracted from a meta-analysis of Patient Health Questionnaire-9 (PHQ-9) scores \citep{levis2019accuracy}, and specifically we focus on the study labeled `Lamers' \citep{lamers2008summed}.

\begin{table}[ht]
\centering
\begin{small}
\begin{tabular}{llccccc lccccc}
\toprule
 &  & \multicolumn{5}{c}{Group 1} & \multicolumn{5}{c}{Group 2} \\
\cmidrule(lr){2-7} \cmidrule(lr){8-13}

& Method & $q$ & mean & sd & SSE & $p$ 
& Method & $q$ & mean & sd & SSE & $p$ \\
\midrule

% Example rows (replace with your data)
\multirow{7}{*}{\rotatebox{90}{Lozovoy}} & Wan 
& $q_1= 2.00 $ & 4.41 & 3.97 &  &  
& Wan & $q_1= 2.00 $ & 5.26 & 4.53 &  &  \\

& BC 
& $m = 3.90$ & 6.16 & 6.56 &  &  
& BC & $m = 5.95 $ & 5.64 & 1.72 &  &  \\

& QE 
& $q_3 = 7.25 $ & 5.99 & 7.14 &  &  
& QE & $q_3 = 7.95 $ & 5.30 & 4.41 &  &  \\

& lognorm 
& $n = 70 $ & 5.99 & 7.14 & 0.004 & 0.810 
& unif & $n = 53 $ & 5.30 & 3.44 & 0.634   & 0.080 \\

& gamma 
&  & 5.24 & 4.57 & 0.014 & 0.698 
& logis & &  5.30 & 4.91 & 0.634 & 0.076 \\

& invGauss 
&  & 5.90 & 6.31 & 0.015 & 0.642 
& norm &  & 5.30 & 4.41  & 0.634 & 0.064 \\

& weibull 
&  & 5.17 & 4.40 & 0.023 & 0.594 
& weibull &  & 5.96 & 4.26 & 1.403 & 0.002 \\ \midrule

% Example rows (replace with your data)
\multirow{7}{*}{\rotatebox{90}{Abo-Shanab}} & Wan 
& $a= 1.13 $ & 36.74 & 39.42 &  &  
& Wan & $a= 1.00 $ & 3.13 & 1.27 &  &  \\

& BC 
& $m = 25.5$ & 38.45 & 43.08 &  &  
& BC & $m = 3.00 $ & 3.18 & 1.30 &  &  \\

& QE 
& $b = 178 $ & 41.63 & 45.85 &  &  
& QE & $b = 6.00 $ & 3.20 & 1.46 &  &  \\

& gamma 
& $n=50$ & 40.93 & 45.95 & 0.624 & 0.382 
& weibull & $n = 25 $ & 3.20 & 1.46 & 0.002   & 0.668 \\

& weibull 
&  & 40.61 & 45.89 & 0.515 & 0.370 
& gamma & &  3.11 & 1.44 & 0.024 & 0.632 \\

& invGauss
&  & 39.48 & 44.72 & 12.9 & 0.050 
& unif &  & 3.33 & 1.57  & 0.167 & 0.306 \\

& lnorm
&  & 44.76 & 48.91 & 110.4 & $<0.001$ 
& logis &  & 3.33 & 1.43 & 0.167 & 0.256 \\ \midrule

% Example rows (replace with your data)
\multirow{7}{*}{\rotatebox{90}{Lamers}} & Wan 
& $q_1= 3.00 $ & 6.76 & 6.77 &  &  
&  &  &  &  &  &  \\

& BC 
& $m = 5$ & 10.52 & 15.56 &  &  
&  &  & &  &  &  \\

& QE 
& $q_3 = 12 $ & 10.64 & 17.80 &  &  
&  &  &  &  &  &  \\

& invGauss 
& $n=104$ & 10.50 & 15.04 & 0.287 & 0.116 
&  & &   &  &  & \\

& lnorm
&  & 10.64 & 17.80 & 0.466 & 0.072 
&  &  &  &  &  & \\

& gamma 
&  & 8.50 & 8.78 & 1.102 & 0.020 
&  &  &  &  &  &  \\

& weibull
&  & 8.58 & 9.08 & 1.07 & 0.016 
&  & &  &  &    &  \\

\bottomrule
\end{tabular}
\end{small}
\caption{Summary of results for mean and SD estimation from three-number summaries using the Wan, BC and QE methods, as well results from parameter estimates arising from fitting to specific distributions. For the specific distributions we present the top four as ranked by having either small SSE or large bootstrap p-values $(p)$.  The three-number summaries are also shown.}
\label{tab:il17_summary}
\end{table}

Recall that the Wan method was used to estimate the mean and SD for the Lozovoy and Abo-Shanab data.  For Lozovoy Group 1, the sensitivity analysis suggests that the Luo/Wan approach is likely to be a large underestimate of both the mean and SD.  The QE method chooses the lognormal distribution, and this distribution had the lowest SSE and the highest boostrap p-value of the distributions we considered.  Other possibilities based on the bootstrapping process are the Gamma, Weibull and inverse-Gaussian, and it can be seen that the SD estimates vary from 4.4 to 7.14.  This should be highlighted as part of the sensitivity analysis.    For Group 2, the tested distributions are close to being rejected (at a level of 5\%) as plausible, and this is likely due to the direction of skew in the three-number summaries not being suited to right-skew fits.  Preference is therefore given to symmetric shapes.  The BC approach had a very small SD compared to the others whose SD estimates varied between 3.44 and 4.91 (not including Weibull, which had a small bootstrap p-value of 0.002).

For Abo-Shanab Group 1, QE chose the Gamma distribution, which also had the smallest SSE for the tested distributions.  Note that the mean and SD are slightly different due to small differences in the optimization of the Gamma parameters.  The `estmeansd' package estimates the shape and rate parameter.  We favored optimizing the shape and scale parameter (inverse of shape).  The clear skew in the three-number summary results in the symmetric distributions being rejected, with the Weibull distribution being another good option. For these data, the estimated SD is similar for both the gamma and Weibull, and it could reported that estimated SDs are reasonably stable for suitable distributions that were tested.  For Group 2, SDs arising from the top distributions varied between 1.43 and 1.57.

For the Lamers data, the QE approach chooses the lognormal and provides an estimated SD of 17.8.  However, the inverse-Gaussian provides a better fit with a reduced SSE and superior bootstrapping results.  The estimated SD is almost 15\% smaller at 15.04 which is enough to have a notable impact on any subsequent inference.    In comparison, the Wan SD is likely to be a vast underestimate, as were the SDs suggested by distributions other than lognormal and inverse-Gaussian.  However, these distributions provided poor fits and were rejected by the bootstrapping process at the 5\% level.  The sensitivity analysis also suggests that choosing an estimated mean of approximately 10.5 is a better choice than the estimated mean of Luo of 6.76.   The impact of a poorly chosen method for estimating the mean and SD can be large.  For example, 95\% confidence intervals (Wald-type: est $\pm 1.96\times$SE) using the Luo/Wan estimates versus the inverse-Gaussian estimates are
$$(5.46,\ 8.06)\;\;\text{and}\;\;(7.61,\ 13.39)$$ are in almost complete disagreement.

\end{example}

\section{Including (cohort) population minimum and maximums as additional information}\label{QE-beta}

In this section, we specifically consider scenario $S_2$ (median and IQR). Although estimators tend to perform best under this scenario compared to $S_1$ (median and range), an additional advantage is that there often exists other information that can be used to improve estimation.  We consider additional information in the form of population (cohort) minimum $(L)$ and maximums $(U)$ that may possibly be available in publications reporting medians and IQRs. We note that this is distinctly different from using sample minimums and sample maximums.  We emphasize that populations may be limited to a specific cohort specific to the study in question.  For example, in a weight loss study while $L=15$ and $U=50$ would cover almost all individuals in a population that does not include extremes, a population minimum of 15 for those targeted in a weight loss trial would not be suitable.   However, inclusion criteria may specify the minimum directly (e.g., \textit{only participants of a BMI greater than 25 were enrolled}).

Some other examples where we may be able to choose and $L$ and/or $U$ to include are:
\begin{description}
    \item[Age (yrs):] Inclusion criteria for studies often include statements such as \textit{all adults over the age of 18 years were included for analysis} $(L=18)$, or \textit{children under the age of 18 were included} ($U=18$ and potentially $L=0$).   It is common to report age as median and IQR instead of mean and SD.
    \item[Length of stay (days):] Length of stay in hospital is often reported as medians instead of means.  For example, in a meta-analysis of length of stay by \cite{sauro2024enhanced}, of the 44 studies reporting a mean or median, more than half reported the median and IQR or range.  A natural population minimum is $L=1$ (same-day discharge) when condition/treatment is non-specific, although a larger $L$ may be necessary for some treatment or medical conditions.
    \item[Patient Health Questionnaire-9 (PHQ-9):] Data from questionnaires that have been aggregated are also often reported as medians and modeled as continuous data.  As an example, the PHQ-9 questionnaire (which we considered in Example \ref{example:real_sensitivity}) consists of 9 questions, each given a value of 0, 1, 2 and 3 and the total is then aggregated ($L=0$ and $U=27$).
\end{description}

In our initial explorations fitting flexible distributions to the median, IQR and $L$ and/or $U$, the results were mixed.  For example, many distributions have infinite $U$ (e.g. lognormal) and infinite $L$ and $U$ (e.g. normal), and so finite $L$ and/or $U$ could only be achieved using truncation and the results were mixed. Instead, we consider flexible fitting of three-number summaries with models that explicitly allow for fixed and finite $L$ and/or $U$.  The two new scenarios that we consider are the following:
\begin{description}
    \item[$S_4=S_2+L$:] scenario $S_2$ ($q_1$, $m$, $q_3$, $n$) with a known $L$ and,
    \item[$S_5=S_2+L+U$:] scenario $S_2$ ($q_1$, $m$, $q_3$, $n$) with a known $L$ and $U$. 
\end{description}
Although we do not explicitly consider it here, it is also possible to consider $S_2+U$.

\subsection{Methods for five-number summaries}

Several existing methods use the five-number summary in scenario $S_3$, where both the sample extremes, minimum ($a$) and maximum ($b$), and the quartiles ($q1, m, q3$) are available. Under an assumption of underlying normality, \cite{wan2014} proposed estimating SD by combining information from both the range and IQR as
\[
\tilde{s}_{3} = \frac{b-a}{4\Phi^{-1}\left(\frac{n-3/8}{n+1/4}\right)} + \frac{q_3-q_1}{4\Phi^{-1}\left(\frac{3n/4-1/8}{n+1/4}\right)}.
\]
An updated version of the above formula was proposed by \cite{shi2020}, who used sample size based weights for the range and IQR components, and this estimate is given as
\[
\tilde{s}_{3} = \frac{b-a}{\left(2+0.14n^{0.6}\right)\Phi^{-1}\left(\frac{n-3/8}{n+1/4}\right)} + 
\frac{q_3-q_1}{\left(2+\frac{2}{0.07n^{0.6}}\right)\Phi^{-1}\left(\frac{3n/4-1/8}{n+1/4}\right)} .
\]
The QE and BC symmetrization approaches can also be applied to the five-number summary \citep{mcgrath2020}. For QE, the quantile matching approach matches $a$ and $b$ to the theoretical $1/n$ and $1-1/n$ quantiles, respectively, so that
\[
\begin{aligned}
\hat{\boldsymbol{\theta}}_{3j}:=\arg\min_{\boldsymbol{\theta}_j}
\Bigl[
&\left(F^{-1}_{j,\boldsymbol{\theta}_j}(1/n)-a\right)^2+ \left(F^{-1}_{j,\boldsymbol{\theta}_j}(1/4)-q_1\right)^2+
\left(F^{-1}_{j,\boldsymbol{\theta}_j}(1/2)-m\right)^2 \\
&+
\left(F^{-1}_{j,\boldsymbol{\theta}_j}(3/4)-q_3\right)^2+\left(F^{-1}_{j,\boldsymbol{\theta}_j}(1-1/n)-b\right)^2
\Bigr],
\end{aligned}
\]
where $F_{j,\boldsymbol{\theta}_j}$ denotes the distribution function of the $j$th candidate distribution (normal, lognormal, gamma, Weibull or beta) with parameter vector $\boldsymbol{\theta}_j$ .

For the BC method, Box-Cox $\lambda$ is chosen such that the transformed quartiles and extremes are approximately symmetric around the transformed median.
\[
\hat{\lambda} = \arg\min_{\lambda} \left\{ 
\bigl[\bigl(t(q_3;\lambda)-t(m;\lambda)\bigr) - \bigl(t(m;\lambda)-t(q_1;\lambda)\bigr)\bigr]^2 +
\bigl[\bigl(t(b;\lambda)-t(m;\lambda)\bigr) - \bigl(t(m;\lambda)-t(a;\lambda)\bigr) \bigr]^2
\right\},
\]
where $t(; \lambda)$ is the Box-Cox transformation function. See Section \ref{sec:methods} for a clearer understanding of these methods. 

Although these five-number summary methods use $a$ and $b$, these are sample minimum and maximum values which can be very different assumed known population bounds $L$ and $U$ in scenarios $S_4$ or $S_5$. The above methods may perform well when $a$ and $b$ are close to $L$ and $U$ (which it could be reasonable to assume for a sufficiently large sample $n$), because the minimum and maximum of the reported sample are estimates of population limits.    However, how large a sample size, $n$, should be to be able to assume that $a$ and $b$ are close to $L$ and $U$ depends on the probability mass in the distributions which is unknown.

\subsection{A new approach based on the Beta distribution: QE-beta}\label{subsec:GBeta}

The Beta distribution defined in $[0,\ 1]$, which is used in the QE approach, is well known for its flexibility in modeling various shapes.  For example, shapes include uniform, bell-shaped, left- and right-skew as well as $U$-shaped and shapes that are useful for describing floor and ceiling effects (e.g. $J$-shapes).  Let $Q(p)$ denote the quantile function for the Beta distribution which is defined in terms of its shape parameters, $\alpha$ and $\beta$.  Then a scaled Beta  (SB)  distribution (or four parameter Beta) can be defined in terms of its quantile function by
\begin{equation}
    Q_{l,u}(p) = L + (U-L)\cdot Q(p) \in [L,\ U].\label{eq:SB}
\end{equation}

When $L$ or $U$ (or both) are known, we propose a variant of the QE method called QE-beta which only fits the scaled Beta for either of the $S_4$ or $S_5$ scenarios.  As noted above, this distribution includes much flexibility and can approximate normal, lognormal, Gamma, Weibull, and many other distributions well.

\begin{figure}[h!t]
    \centering
    \includegraphics[width=0.75\linewidth]{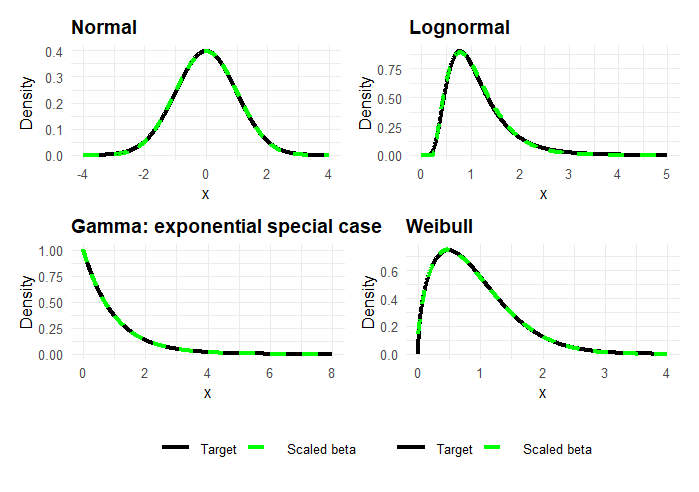}
    \caption{Probability density function plots (black curve) and approximating scaled Beta distribution (yellow dashed line) for the $N(0,1)$, LogN$(0, 1)$, exponential and Weibull$(1.5, 1)$ distributions.}
    \label{fig:SB_approxs}
\end{figure}

We plot the probability density functions for the normal, lognormal, exponential, and Weibull distributions in Figure \ref{fig:SB_approxs}.  By optimizing an objective function as the sum of squared differences between the SB and target distribution, we were able to determine parameters for which the SB distribution approximated the target very well.  Hence, the SB distribution is capable of approximating many shapes well, and if $L$ and $U$ are known, then only two parameters need to be estimated.

Fixing $L$ and $U$, we estimate the Beta shape parameters, $\alpha$ and $\beta$, using the three-number summary.  The easiest way to do this is to transform $q_1$, $m$ and $q_3$ into
\begin{equation}
    \widetilde{q}_1=\frac{q_1-L}{U-L},\;\widetilde{m}=\frac{m-L}{U-L},\;\widetilde{q}_3=\frac{q_3-L}{U-L}
\end{equation}
and estimate the shape parameters for the standard Beta distribution using $\{\widetilde{q}_1, \widetilde{m}, \widetilde{q}_3\}$ and $n$.  To do so, we used the same QE objective function shown in Section \ref{section:QE} with the candidate distribution restricted to Beta only.

When only $L$ is known, we modify the QE objective function to include the unknown $U$ as the third parameter. Although this increases the dangers of overfitting, our simulation results shown later suggest the approach can work well, as much of the shape information from the SB can be obtained from the spacings in $\{L,q_1,m,q_3\}$.

\section{Simulations and examples}

\subsection {Simulations using generated data}\label{sec:simu}

We conducted a simulation study to evaluate the performance of sample SD estimation methods when additional information about the minimum and maximum population is available alongside the three-number summary $\{q_1,m,q_3\}$. Specifically, the aim of the simulations is (i) to compare the performance of the proposed method with existing SD estimation methods in settings where external lower and upper bounds (population minimum and maximum) are available and (ii) to examine the corresponding four-number settings where only a lower bound (population minimum) is available. 

\subsubsection {Simulation parameters and performance measures}

%This simulation setup reflects applications where possible global minimum and/or maximum are known or can be justified from external research, but the observed sample minimum and maximum would not necessarily be known. For example, if the outcome is age in adults, a lower bound such as 18 years and an upper bound such as 120 years may be known even when the sampled minimum and maximums are unknown. 

To represent a wide range of symmetric, skewed, heavy-tailed, and non-standard shapes, we considered various parameter combinations of the SB distribution with lower bound $L=18$ and upper bound $U=120$. The following parameter combinations were used; $(\text{shape}_1,\text{shape}_2)\in\{(1,1),\,(0.5,0.5),\,(2,8),\,(8,2),\,(2,2),\,(5,5),\,(0.8,0.2),\,(0.2,0.8)\}$. These parameter settings correspond to distributions with different shapes, namely uniform, U-shaped, right-skewed, left-skewed, symmetric with a moderate central peak, symmetric with a strong central peak, highly left-skewed/J-shaped, and highly right-skewed/reverse J-shaped, respectively. The probability densities for all distributions considered are shown in Figure \ref{Simu_shapes} of the Supplementary Information.  

We considered sample sizes of $n = 30,\ 50,\ 100,\ 200,\ 500,\ 1000$ and $1000$ trials to be carried out for each distribution and $n$. For each trial, the sample quartiles $q_1, m, q_3$ were calculated, together with the true SD sample, denoted by $s$. The sample quartiles and median formed the  three-number summary, while the population lower bound ($L=18$) and the upper bound ($U=120$) were used to estimate the SD using the proposed five- and four-number procedures based on the QE-beta approach.  We computed the relative error (RE) between each estimated SD ($\tilde{s}$) and the true sample SD, $s$, from each trial as
\[
\mathrm{RE}_{\tilde{s}}
=
\frac{\tilde{s} - s}{s}\times 100\%.
\]
The average RE was then used to assess the performance of the SD estimators.

\subsubsection {Results of estimating SD with a known L and U}\label{subsec:simu_5p}

For the five-number summary scenario (where $L$ and $U$ are the know population minimum and maximum), we compared the QE-beta estimator with the existing methods of the QE and BC \citep{mcgrath2019estmeansd}, Wan \citep{wan2014} and Shi \citep{shi2020} estimators based on the five-number summary.  These existing methods assume that the given $L$ and $U$ are actually the sample minimum and maximum.  To distinguish these methods from the versions that use the three-number summary, we denote these as in the tables, figures, and text as $\text{QE\_5}$, $\text{BC\_5}$ $\text{Wan\_5}$, $\text{Shi\_5}$ methods .

\begin{figure}[ht]
\centering
\includegraphics[width=0.75\linewidth]{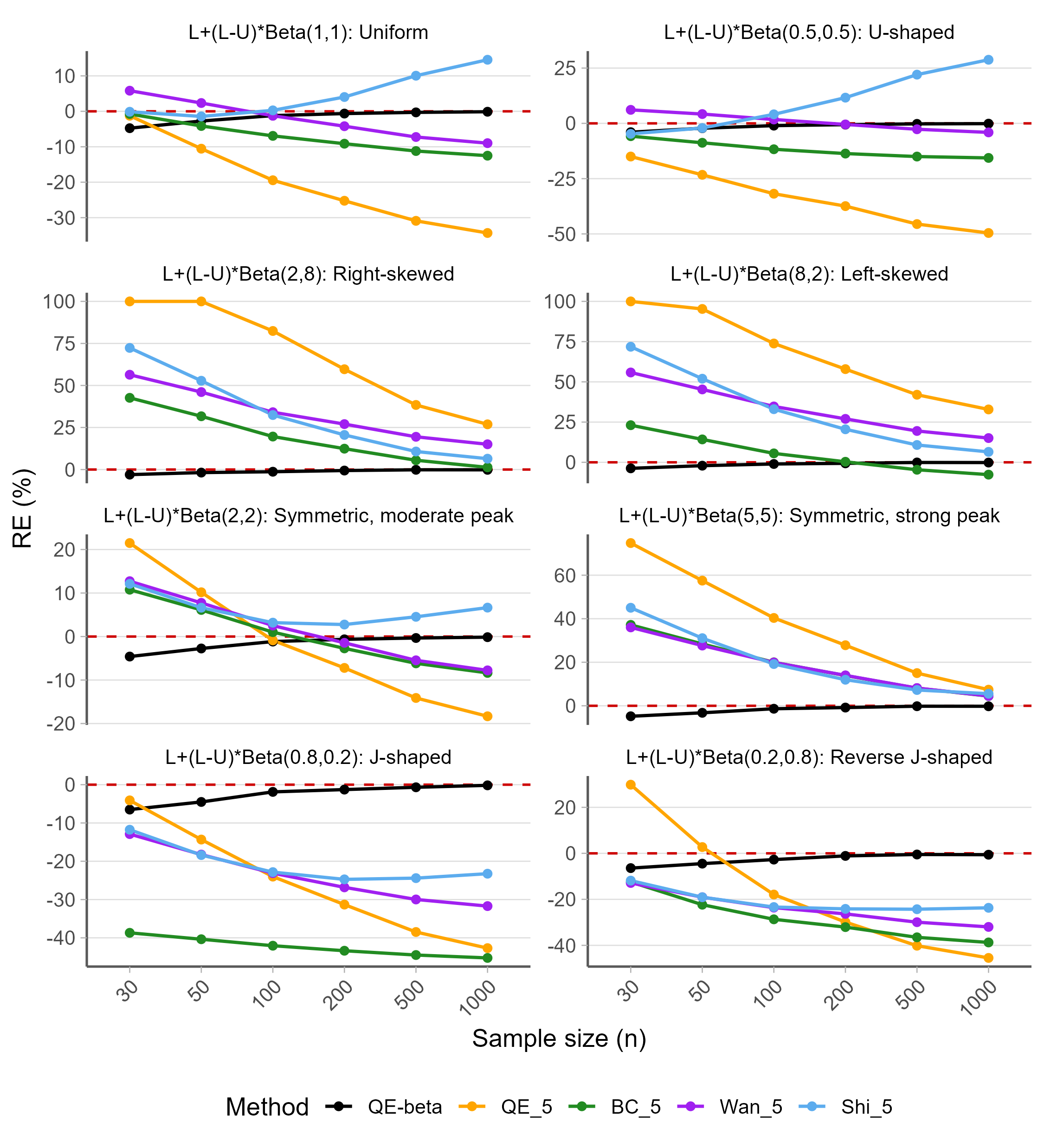}
\caption{\label{fig:Simu_RE_5p} Percentage relative errors of estimating SD from quantiles with known lower (L) and upper (U) bounds using QE-beta and existing methods.}
\end{figure}

Figure~\ref{fig:Simu_RE_5p} presents the REs of SD estimation using the methods described above for the distributions considered.  Overall, the proposed QE-beta method performs well in all the shapes and sample sizes considered.  The REs are consistently close to zero and generally improve as the sample size increases.  This behavior is particularly clear for the skewed, strongly peaked, and J-shaped distributions where the estimator remains stable. 

 Performance of existing methods assuming that $L$ and $U$ are the sample extremes are mixed.  The largest REs are observed for the skewed SB distributions with parameters (2, 8) and (8, 2). In particular, \text{QE\_5} substantially overestimates the SD under these distributions, with REs exceeding 100\% for smaller sample sizes. Similar patterns are observed for the strongly peaked symmetric distribution, scaled Beta(5, 5). The \text{BC\_5}, \text{Wan\_5} and \text{Shi\_5} methods also tend to overestimate the SD in these settings, although generally to a lesser extent than \text{QE\_5}. Under the J-shaped and reverse J-shaped distributions, most existing methods show underestimation, particularly for moderate and larger sample sizes.   Methods such as \text{QE\_5}, \text{BC\_5}, \text{Wan\_5} and \text{Shi\_5} were developed under the assumption that the minimum and maximum arise directly from the observed sample and therefore contain information about the sample spread through the sample size.  For some of the distributions (right-skewed, left-skewed and symmetric), the performance of the existing methods improves as $n$ increases. This occurs because the sample minimums and maximums are likely to be close to the assumed known $L$ and $U$.  For the other scenarios, an improvement is not observed, mainly due to large deviations away from the assumed underlying normality.  These results illustrate that existing five-number summary methods may or may not be well-suited to settings where externally specified population bounds are supplied in place of observed sample minima and maxima. In contrast, the QE-beta approach works well for all of the shapes considered.   The corresponding numerical values of the REs of this section are provided in the Supplementary Table \ref{tab:gbeta_5p}.

\subsubsection {Results of estimating SD with only a known L}\label{subsec:simu_3p_vs_4p}

The second part of the simulations examined the proposed QE-beta method for estimating SD using a four-number summary when a population lower bound (taken here to be $L=18$) is available. Unlike the five-number setting, there are no widely used existing methods designed specifically for such four-number summaries. Therefore, we compare the QE-beta with the existing three-number summary methods denoted $\text{QE\_3}$ and $\text{BC\_3}$ \citep{mcgrath2019estmeansd}, and $\text{Wan\_3}$ \citep{wan2014}.  In the QE-beta method, the lower bound $L$ is treated as fixed and known, while $U$ needs to be estimated.

\begin{figure}[ht]
\centering
\includegraphics[width=0.75\linewidth]{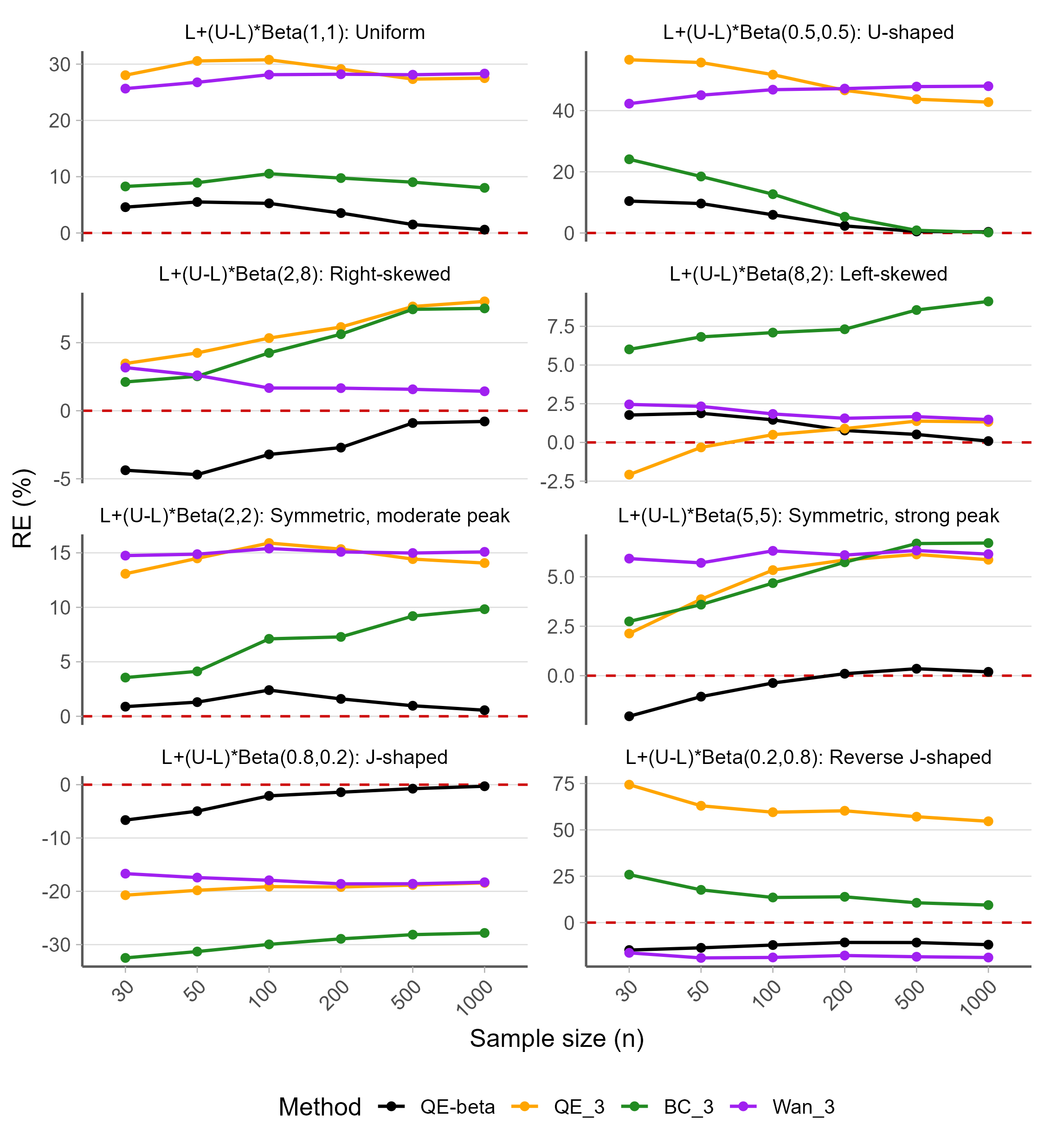}
\caption{\label{fig:Simu_RE_4p} Percentage relative errors of estimating SD from quantiles with a known lower bound (L) using QE-beta compared with existing three-number summary methods.}
\end{figure}

Figure~\ref{fig:Simu_RE_4p} presents the REs. Overall, the proposed QE-method generally produced smaller REs than the three-number summary methods for most distributions and sample sizes. This is clear for the uniform, U-shaped, symmetric, and strongly skewed distributions, where the REs of QE-beta are consistently closer to zero.

The existing three-number summary methods often show overestimation or underestimation depending on the distributional shape. For example, $\text{QE\_3}$ shows a notable overestimation for the uniform, U-shaped, symmetric, and reverse J-shaped scaled beta distributions, while both $\text{BC\_3}$ and $\text{Wan\_3}$ also display a large positive bias in several settings. Under the highly skewed J-shaped distribution, all methods underestimate the SD, although the QE-beta with a known minimum L remains closer to zero than using three-number summaries alone. These are good examples that show that using a meaningful lower population bound alongside the quartiles can provide useful additional information about the support and shape of the underlying distribution. Even when the upper bound is unknown and must be estimated, the QE-beta approach was able to use this additional information to improve SD estimation. The numerical values of the REs of this section can be found in Supplementary Table \ref{tab:gbeta_4p}.

\subsection {Simulations using real data}

To further evaluate the methods using simulations based on a real data example; namely the publicly available PHQ-9 depression assessment dataset by \cite{sebastian_burchert_2019_3384860} that we obtained from the Kaggle website (\url{https://www.kaggle.com/datasets/thedevastator/phq-9-depression-assessment/data}). The data set contains information on the responses to the Patient Health Questionnaire (PHQ-9), consisting of nine items measuring the severity of depressive symptoms experienced during the previous two weeks in a sample of the general population.  We extracted the nine PHQ-9 item responses for each participant and calculated a total PHQ-9 score by summing the 9 item responses. A PHQ-9 item is scored on a four-number ordinal scale ranging from 0 to 3. Consequently, the total PHQ-9 score is naturally bounded between $L = 0$ and $U = 27$ inclusive. The resulting total scores showed only mild skewness with the observed minimum and maximum of 3 and 27, respectively, as shown in Figure \ref{fig:phq9_hist} of the Supplementary Information.

Using the total scores observed for PHQ-9 as an empirical population, we repeatedly sampled large datasets $n = 20, 50, 100, 250$. For each sample size, 100,000 trials were conducted, with or without replacement depending on the sample size. For every simulated sample, the true sample standard deviation, $s$, together with the quantile summaries $\{q_1,m,q_3,n\}$ were calculated to compare the SD estimation approaches. We first used these three-number summaries to estimate the SD using existing methods described in \ref{subsec:simu_3p_vs_4p}. Then we used $\{L,q_1,m,q_3,U,n\}$ to estimate SD using existing five-number summary methods (see \ref{subsec:simu_5p}), as well as the new QE-beta method, where population limits were fixed at $L=0$ and $U=27$. Similarly to the previous section, for each method, we calculated the percentage relative error (RE) between the true sample SD and the estimated SD. 

\begin{figure}[ht]
\centering
\includegraphics[width=0.85\linewidth]{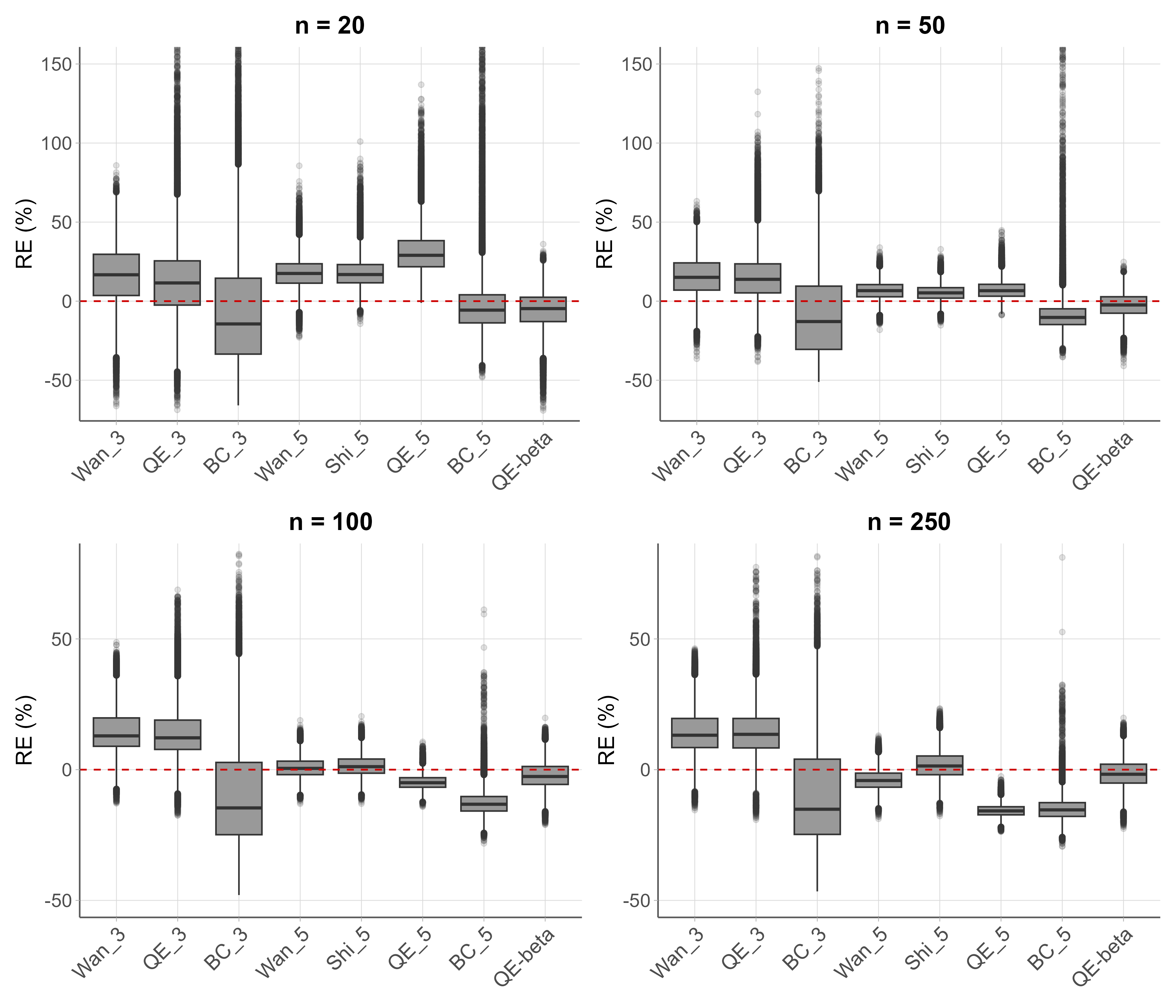}
\caption{\label{fig:phq9_RE} Percentage relative errors for estimating the standard deviation (SD) from PHQ-9 data using QE-beta and existing methods.}
\end{figure}

Figure~\ref{fig:phq9_RE} presents the distributions of the percentage relative errors in the simulation replicates for each method and for each sample size. It shows that using only a three-number summary, the existing methods performed poorly to estimate the SD of the total score of PHQ-9. Across all sample sizes, the $\text{Wan\_3}$, $\text{QE\_3}$ and $\text{BC\_3}$ methods had large bias and variability, where $\text{Wan\_3}$ and $\text{QE\_3}$ severely overestimated the SD and $\text{BC\_3}$ underestimated with greater instability. However, when the known lower and upper bounds were provided through the five-number summary approaches, performance improved generally.

For the smaller sample sizes ($n=20$ and $50$), the QE-beta clearly outperformed the existing five-number summary methods which treat $L$ and $U$ as sample extremes. The $\text{Wan\_5}$, $\text{Shi\_5}$ and particularly $\text{QE\_5}$ methods resulted in an overestimation of SD, with many outlying SDs when $n=20$. Although outliers were present in QE-beta, those were generally limited to approximately 50-60\%, while existing methods like $\text{BC\_5}$ frequently produced many large outliers extending well beyond $100\%$. For $n=100$ the $\text{Wan\_5}$, $\text{Shi\_5}$ and QE-beta methods performed best overall, likely because the observed sample minimum and maximum become closer to the true population limits as the sample size increases, making methods based on sample extremes more appropriate.  However, both $\text{QE\_5}$ and $\text{BC\_5}$ continued to show a negative bias. For $n=250$, where the sampling was performed with replacement as the sample size exceeded the number of individuals in the original data set $(N=189)$, the QE-beta again showed good results in estimating SD with only slight underestimation. 

\begin{figure}[ht]
\centering
\includegraphics[width=0.9\linewidth]{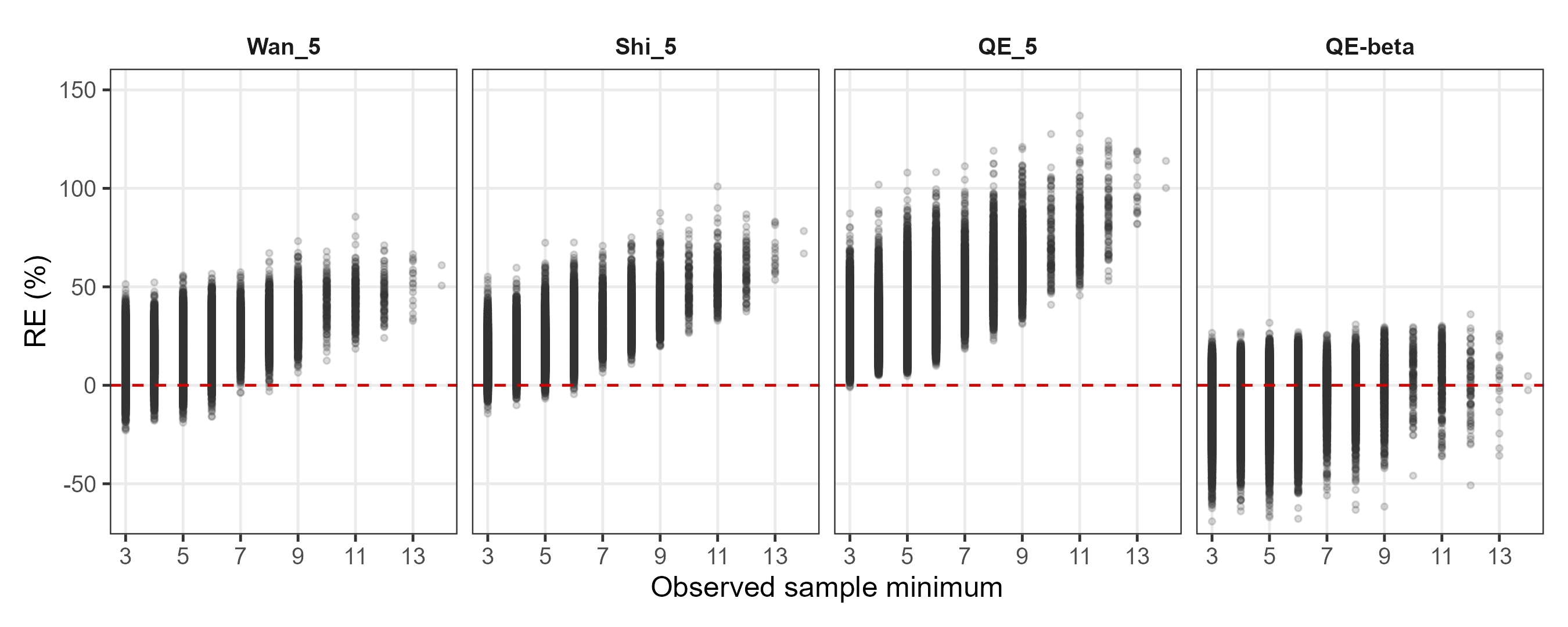}
\caption{\label{fig:phq9_RE_vs_min} Percentage relative errors of estimating SD versus observed sample minimums of PHQ9 samples.}
\end{figure}

To better understand these deviations observed when $n=20$, we additionally plotted the REs against the observed sample minimum of simulated samples in Figure \ref{fig:phq9_RE_vs_min}. Although the overall observed minimum of the total PHQ-9 was 3, the assumed lower bound was fixed at $L=0$, corresponding to the theoretically possible minimum. For the $\text{Wan\_5}$, $\text{Shi\_5}$ and $\text{QE\_5}$ methods, which treat the supplied lower bound as if it were the observed sample minimum, the REs became increasingly large (showing a upward trend) as the observed sample minimum moved further away from the assumed $L=0$. In contrast, the QE-beta remained comparatively stable throughout the range of observed sample minimum values, likely because $L=0$ was treated as a fixed population limit rather than as a sample minimum. 

In the Supplementary Information, we provide several other figures that are useful for assessing the performance of the methods.  In Figure \ref{RE_vs_BS}, we show the RE versus the Bowley skew measure computed for each sample and for varying $n$.  We can see that the QE and BC methods are most sensitive to skewness for three-number summaries.  For the five-number summary with $L$ and $U$, the BC approach returned poor results for increasing positive skew where the other methods were relatively stable.  The Wan, Shi, and QE methods tended towards overestimation for small to moderate sample sizes, while QE-beta tended towards underestimation for small to moderate sample sizes. In Figure \ref{RE_vs_BS_ran_n} we display these results for a random $n$ and these relationships become clearer. The box plots of the RE for randomly chosen $n$ are shown in Figure \ref{RE_ran_n}, highlighting the good performance of Wan, Shi, and QE-beta for the five-number summary, where the poor results of Wan tend to be overestimation, and the poor results of QE-beta tend to be underestimation. Figure \ref{RE_vsn_ran_n} shows that for small sample sizes, a very large over-estimation can occur for the Shi and QE methods. 

These real-data applications support the findings from the generated data simulations that using meaningful population bounds through the scaled beta distribution tend to improve SD estimation when only quantile summaries are available. See the supplementary information section \ref{section:phq9_more} for more details and the results of these real-data simulations.

\section{Summary and recommendations}

Our explorations highlight that attempting to estimate SD from three-number summaries is inherent with risk.  There is not enough information to determine which method is suitable or whether the result is even moderately close to the unknown sample standard deviations. Even for symmetric three-number summaries, methods that assume an underlying normal distribution can have a very large error in estimation with normality not truly evident.  Poor estimation can lead to greatly overestimated effects and invalid inference.  We therefore have the following recommendations.

\begin{enumerate}
    \item If a meta-analysis is intended, instead meta-analyze medians.  In recent years, several methods and related R packages have been introduced for the meta-analysis of medians and exhibit good inference properties \citep[e.g.][]{mcgrath2019one, mcgrath2020meta, metamedian, metaquant, de2024novel}. Although separate meta-analysis of means and medians may reduce power, it may be possible to use a meta-regression analysis to model them together. 
    \item If an estimated SD is required, perform a sensitivity analysis.  It is good practice to perform sensitivity analyses when the validity of the assumptions required is not known \citep[e.g.][]{Design_of_Obs_studies}.  We recommend using several methods such as those of \cite{wan2014} and \cite{mcgrath2020} as well as estimating SD assuming a variety of distributional shapes to understand how much the estimated SD may vary from the target.  We have introduced a bootstrap goodness-of-fit test that allows some shapes to be deemed inappropriate and allowing focus on shapes that are plausible.  We recommend that researchers describe the results of the sensitivity analysis and discuss the impact variation in the estimated SDs can have on any analysis that follows.
    \item Seek additional information.  In some cases, a population minimum and/or maximum may be known.  When sample sizes are small, we found that a new approach based on the scaled Beta distribution with known minimum and maximum can perform well due to shape flexibility.  For larger sample sizes, existing methods of \cite{wan2014} and \cite{shi2020} that require sample minimum and maximums can perform well when the skewness is not large or when difficult shapes are evident (e.g., shapes $J$- and $U$-). 
\end{enumerate}

We believe that utilizing at least one of the above recommendations will improve statistical rigor and transparency and allow researchers to have more confidence in their findings or be able to discuss potential biases. 

\section{A Shiny web application for sensitivity}

We have created a Shiny \citep{shiny} we application that allows users to perform the sensivity analyses for various methods and models when analyzing three-number summaries.  The web application can be found here:

\begin{center}
    \url{https://mathstats-drama.shinyapps.io/mean_sd_sensitivity/}
\end{center}

The key features of the application are:
\begin{itemize}
    \item Three-number summary estimates of the mean and SD are obtained using the Luo/Wan, QE, and BC approaches, as well as for several probability distributions.
    \item Three-number summaries can be entered in the form of scenarios $S_1$ or $S_2$.
    \item The application can also perform the bootstrap goodness-of-fit test so that users can rule out some data shapes and the resulting estimates to focus on those that are more plausible.
    \item Users can download the results.
\end{itemize}

A \textit{User Guide} instruction tab is included.  Since the app will continue to be improved over time (e.g. new method, distributions etc.), we direct the read to this tab for more info.

%\section{Conclusion}
%Your conclusion here

%\section*{Acknowledgments}
%This was was supported in part by......

%Bibliography

\bibliographystyle{authordate4}
\bibliography{references}

% Appendix
\begin{appendix}
\section{Supplementary Information}\label{appendixA}
\setcounter{figure}{0}   
\setcounter{table}{0}
\renewcommand{\thefigure}{S\arabic{figure}}
\renewcommand{\thetable}{S\arabic{table}}

\subsection{Details of the optimization process using in the sensitivity analysis}\label{section:optim}

In this section, we provide details of the starting values of the parameters for the optimization process to minimize the SSE of the sample and the theoretical quantiles.  First, we define Bowley's quantile skew measure which is useful for choosing as starting value for the Gamma distribution shape, and also as to whether it is suitable to fit the Pareto distribution to the sample quartiles.  It is defined as
\begin{equation}
    B = \frac{q_1+q_3 - 2m}{q_3-q_1}\label{eq:bowleys_skew}
\end{equation}
and is widely considered as the preferred robust alternative to the moment-based measure of skew.

\begin{table}[ht]
    \centering
    \begin{small}
\begin{tabular}{llll}
\toprule
    Distribution & starting values & mean & SD \\ \midrule
     $N(\mu, \sigma^2)$ & $\widehat{\mu}_0=m$,\;\; $\displaystyle\widehat{\sigma}_0 = \frac{q_3-q_1}{2\sqrt{2}\cdot\text{erf}^{-1}(0.5)}$ & $\widehat{\mu}$ & $\widehat{\sigma}$\\
     Unif$(a,b)$ & $\widehat{a}_0=2q_1-m$,\;\;$\widehat{b}_0=2q_3-m$ & $\displaystyle\frac{\widehat{a}+\widehat{b}}{2}$ & $\displaystyle\frac{\widehat{b}-\widehat{a}}{\sqrt{12}}$\\ 
     Logis$(\mu,s)$ & $\widehat{\mu}_0=m$,\;\;$\displaystyle\widehat{s}_0 = \frac{q_3-q_1}{2\text{log}(3)}$ & $\widehat{\mu}$ & $\displaystyle\frac{\widehat{s}\pi}{\sqrt{3}}$\\
     LN$(\mu,\sigma)$ & $\widehat{\mu}_0=\text{log}(m)$,\;\; $\displaystyle\widehat{\sigma}_0 = \frac{\text{log}(q_3)-\text{log}(q_1)}{2\sqrt{2}\cdot\text{erf}^{-1}(0.5)}$ & $\exp\left(\widehat{\mu}+\frac{\widehat{\sigma}^2}{2}\right)$& $\sqrt{\left[\exp(\widehat{\sigma}^2) - 1\right]\exp(2\widehat{\mu}+\widehat{\sigma}^2)}$\\
     Gamma$(\alpha,\theta)$ & $\displaystyle\widehat{\alpha}_0=\left(\frac{1}{4B}\right)^2$,\;\; $\displaystyle\widehat{\theta}_0 = \frac{m}{Q(0.5;\widehat{\alpha}_0,1)}$ & $\widehat{\alpha}\widehat{\theta}$ & $\sqrt{\widehat{\alpha}}\widehat{\theta}$ \\
     Weibull$(k,\lambda)$ & $\displaystyle\widehat{k}_0=\frac{\text{log}(2)}{\text{log}(q_3/m)}$,\;\; $\displaystyle\widehat{\lambda}_0 = \frac{m}{\left[\text{log}(2)\right]^{1/\widehat{k}}}$ &$\widehat{\lambda}\Gamma(1+1/\widehat{k})$ &$\widehat{\lambda}\sqrt{\Gamma(1+2/\widehat{k})-\Gamma(1+1/\widehat{k})^2}$\\
     Pareto$(x_{\text{min}},\alpha)$ & $\hat x_{\text{min},0}=\displaystyle 2^{\displaystyle\text{log}(q_1/m)/\text{log(3/2)}}\cdot m$ & $ \displaystyle\frac{\widehat{\alpha}\cdot\hat x_{\text{min}}}{\widehat{\alpha}-1}
,\ \widehat{\alpha}>1$ & $ \displaystyle\frac{\sqrt{\widehat{\alpha}}\cdot\hat x_{\text{min}}}{(\widehat{\alpha}-1)\sqrt{\widehat{\alpha}-2}}
,\ \widehat{\alpha}>2$ \\
     & $\widehat{\alpha}_0 = \displaystyle\text{max}\left\{1, -\frac{\text{log}(3/2)}{\text{log}(q_1/m)}\right\}$ & & \\
     inv-Gauss$(\mu,\lambda)$ & $\widehat{\mu}_0=m$, $\widehat{\lambda}_0 = \displaystyle m\left(\frac{3}{10B}\right)^2$ & $\widehat{\mu}$ & $\displaystyle\sqrt{\frac{\widehat{\mu}^3}{\widehat{\lambda}}}$ \\
     log-Logis$(\alpha, \beta)$ & $\widehat{\alpha}_0 = m$, $\widehat{\beta}_0=2\text{log}(3)/\text{log}(q_3/q_1)$ & $\displaystyle\widehat{\alpha}\cdot \frac{\pi/\widehat{\beta}}{\sin(\pi/\widehat{\beta})}$& $\displaystyle\widehat{\alpha} \cdot \sqrt{
\frac{2\pi/\widehat{\beta}}
{\sin\!\left(2\pi/\widehat{\beta}\right)}
-
\left(
\frac{\pi/\widehat{\beta}}
{\sin\!\left(\pi/\widehat{\beta}\right)}
\right)^2
}$ \\
     \bottomrule
\end{tabular}
\end{small}
    \caption{Starting values for the numerical optimization fitting several distributions to observed sample quartiles and the estimated mean and SD based on the final estimates; $B$ is Bowley's skew shown in \eqref{eq:bowleys_skew} and $Q$ is the quantile function for the Gamma distribution.}
    \label{tab:starting_values}
\end{table}

The starting values for the optimization are shown in Table \ref{tab:starting_values}.  For the Gamma distribution, 
the moment-based skew is equal to $\gamma = 2/\sqrt{\alpha}$. Although $B$ and $\gamma$ can be very different ($B$ is bounded on $[-1,\ 1]$ while $\gamma$ is not), across a wide range of shape $\alpha$ values, an approximation that should provide reasonable starting value choices is $\gamma \approx 8\cdot B$ which gives the starting value $\widehat{\alpha}_0=(1/4B)^2$.  The Gamma scale, $\theta$, follows simply by noting that the ratio of quantile functions with the same shape parameters is equal to the ratio of scale parameters.  For the inverse Gaussian that does not have a closed form for the quantile function, we start with $\widehat{\mu}_0=m$. Although $\mu$ is equal to the mean and will be greater than the median, the starting value should be sufficient for optimization.  Again we use Bowley's skew to find a reasonable approximation to the shape parameter by assuming $\gamma \approx 10B$.

All other starting values were derived directly from the theoretical quantile functions and by replacing the theoretical quartiles with the observed sample quartiles. If optimization fails, we return the starting values as the chosen estimates.  If the starting values do not provide a good fit for the quartiles for the specified distribution, then this will be reflected in a large SSE and likely small bootstrap goodness of fit p-value.

\subsection{More information of simulations}\label{section:shapes}

\begin{figure}[H]
    \centering
    \includegraphics[width=0.75\linewidth]{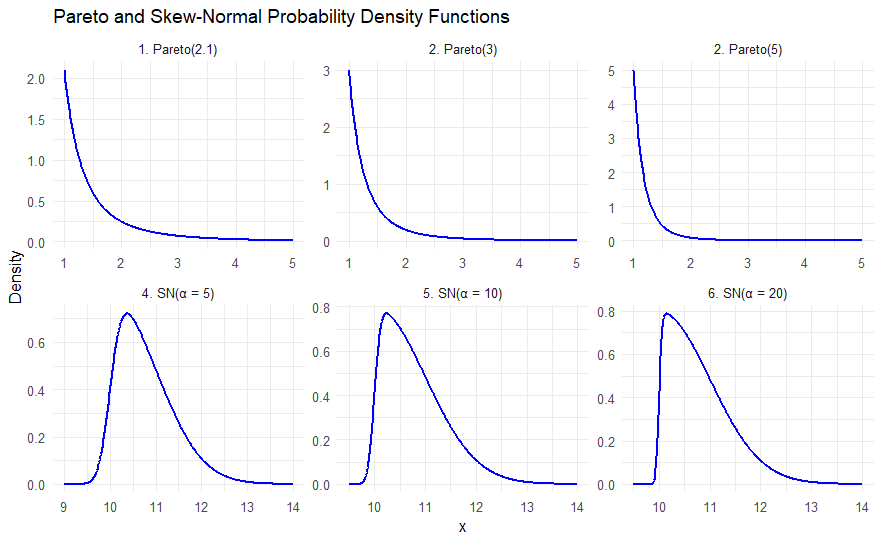}
    \caption{Probability density functions from various Pareto and Skew-Normal distributions with varying slant parameter, $\alpha$.}
    \label{fig:par_and_sk_dist}
\end{figure}

\begin{figure}[H]
\centering
\includegraphics[width=0.8\linewidth]{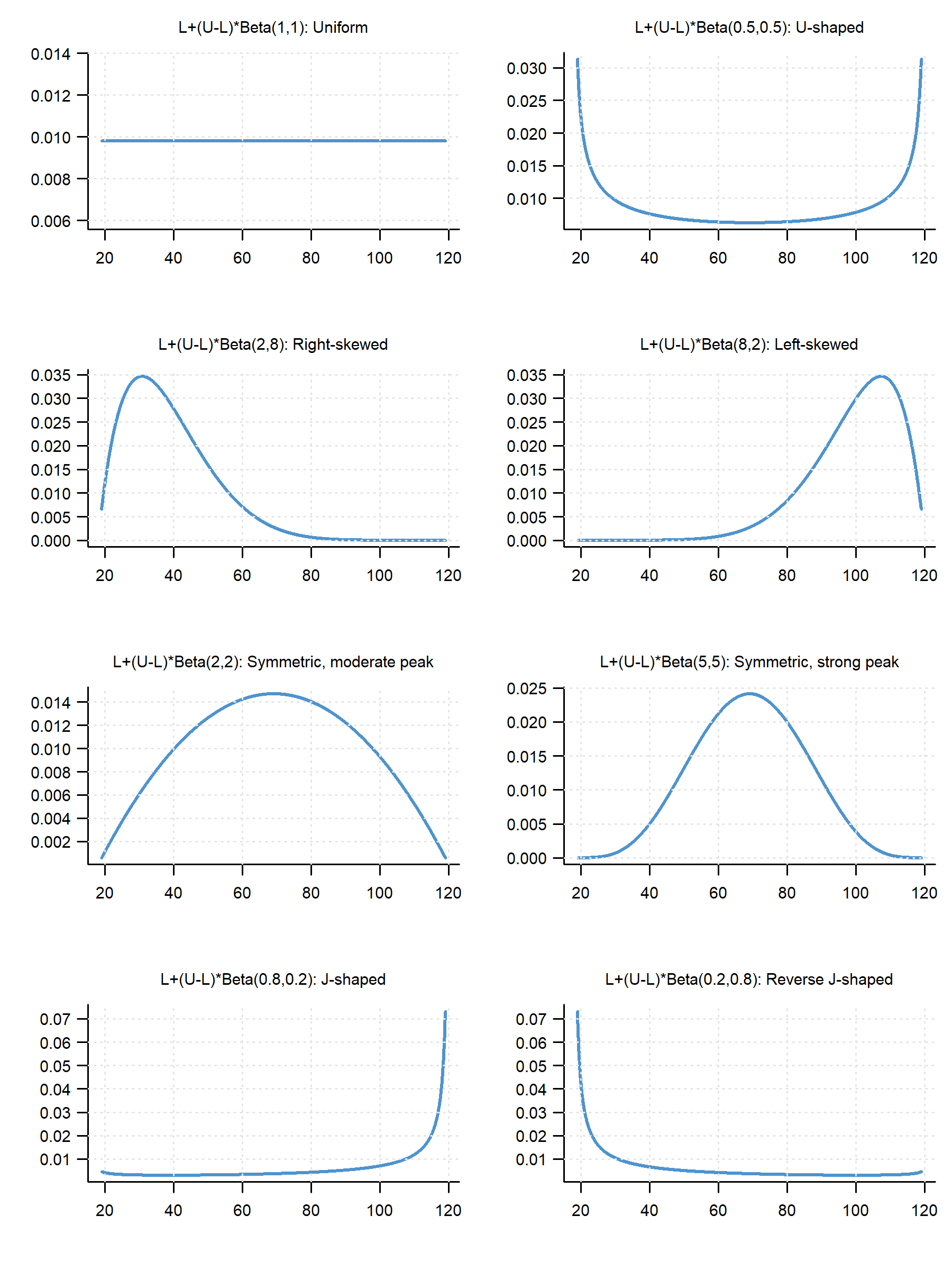}
\caption{\label{Simu_shapes} Shapes of scaled beta distributions used in simulations varying two shape parameters in $L + (U-L)\times \text{Beta}(\text{shape}_1,\text{shape}_2)$, with fixed $L=18$ and $U=120$.}
\end{figure}

\begin{table}[H]
\centering
\caption{Percentage relative errors of estimating SD from quantiles with fixed population minimum and maximum using different shapes of scaled beta distributions.The smallest absolute error in each row is highlighted in yellow and the second smallest is highlighted in green. Note: Scaled Beta $= L + (U-L)\times \text{Beta}(\text{shape}_1,\text{shape}_2)$ with fixed $L=18$ and $U=120$.}
\label{tab:gbeta_5p}
\setlength{\tabcolsep}{6pt}
\begin{tabular}{@{}llrrrrrr@{}}
\toprule
Distribution & Shape remark & $n$ & QE-beta & QE\_5 & BC\_5 & Wan\_5 & Shi\_5 \\
\midrule
\multirow{6}{*}{Scaled Beta $(1,1)$}
& \multirow{6}{*}{Uniform}
& 30   & -4.76 & -1.24 & \cellcolor{secondgreen}-0.76 & 5.82 & \cellcolor{bestyellow}-0.13 \\
& & 50   & -2.74 & -10.54 & -4.14 & \cellcolor{secondgreen}2.34 & \cellcolor{bestyellow}-1.44 \\
& & 100  & \cellcolor{secondgreen}-1.21 & -19.45 & -6.93 & -1.25 & \cellcolor{bestyellow}0.27 \\
& & 200  & \cellcolor{bestyellow}-0.64 & -25.24 & -9.14 & -4.20 & \cellcolor{secondgreen}4.04 \\
& & 500  & \cellcolor{bestyellow}-0.30 & -30.90 & -11.20 & \cellcolor{secondgreen}-7.25 & 10.05 \\
& & 1000 & \cellcolor{bestyellow}-0.12 & -34.33 & -12.52 & \cellcolor{secondgreen}-8.98 & 14.55 \\
\midrule
\multirow{6}{*}{Scaled Beta $(0.5,0.5)$}
& \multirow{6}{*}{U-shaped}
& 30   & \cellcolor{bestyellow}-4.00 & -15.00 & -5.80 & 6.11 & \cellcolor{secondgreen}-4.73 \\
& & 50   & \cellcolor{secondgreen}-2.19 & -23.27 & -8.78 & 4.21 & \cellcolor{bestyellow}-2.11 \\
& & 100  & \cellcolor{bestyellow}-1.03 & -31.83 & -11.71 & \cellcolor{secondgreen}1.75 & 4.09 \\
& & 200  & \cellcolor{secondgreen}-0.58 & -37.42 & -13.66 & \cellcolor{bestyellow}-0.49 & 11.63 \\
& & 500  & \cellcolor{bestyellow}-0.23 & -45.55 & -15.01 & \cellcolor{secondgreen}-2.67 & 22.04 \\
& & 1000 & \cellcolor{bestyellow}-0.13 & -49.58 & -15.63 & \cellcolor{secondgreen}-4.08 & 28.77 \\
\midrule
\multirow{6}{*}{Scaled Beta $(2,8)$}
& \multirow{6}{*}{Right-skewed}
& 30   & \cellcolor{bestyellow}-3.03 & 166.45 & \cellcolor{secondgreen}42.64 & 56.38 & 72.34 \\
& & 50   & \cellcolor{bestyellow}-1.82 & 122.39 & \cellcolor{secondgreen}31.68 & 46.07 & 52.80 \\
& & 100  & \cellcolor{bestyellow}-1.31 & 82.41 & \cellcolor{secondgreen}19.62 & 34.09 & 32.41 \\
& & 200  & \cellcolor{bestyellow}-0.61 & 59.65 & \cellcolor{secondgreen}12.41 & 27.02 & 20.57 \\
& & 500  & \cellcolor{bestyellow}-0.16 & 38.41 & \cellcolor{secondgreen}5.52 & 19.47 & 10.72 \\
& & 1000 & \cellcolor{bestyellow}-0.18 & 26.88 & \cellcolor{secondgreen}1.30 & 15.02 & 6.45 \\
\midrule
\multirow{6}{*}{Scaled Beta $(8,2)$}
& \multirow{6}{*}{Left-skewed}
& 30   & \cellcolor{bestyellow}-3.74 & 116.64 & \cellcolor{secondgreen}23.07 & 55.83 & 71.83 \\
& & 50   & \cellcolor{bestyellow}-2.09 & 95.31 & \cellcolor{secondgreen}14.24 & 45.30 & 51.95 \\
& & 100  & \cellcolor{bestyellow}-1.04 & 73.84 & \cellcolor{secondgreen}5.55 & 34.73 & 33.02 \\
& & 200  & \cellcolor{secondgreen}-0.64 & 57.92 & \cellcolor{bestyellow}0.33 & 26.99 & 20.53 \\
& & 500  & \cellcolor{bestyellow}-0.10 & 42.00 & \cellcolor{secondgreen}-4.62 & 19.48 & 10.77 \\
& & 1000 & \cellcolor{bestyellow}-0.15 & 32.86 & \cellcolor{secondgreen}-7.65 & 15.07 & 6.49 \\
\midrule
\multirow{6}{*}{Scaled Beta $(2,2)$}
& \multirow{6}{*}{Symmetric, moderate peak}
& 30   & \cellcolor{bestyellow}-4.60 & 21.49 & \cellcolor{secondgreen}10.77 & 12.72 & 12.12 \\
& & 50   & \cellcolor{bestyellow}-2.75 & 10.17 & \cellcolor{secondgreen}6.11 & 7.74 & 6.63 \\
& & 100  & -1.16 & \cellcolor{bestyellow}-0.88 & \cellcolor{secondgreen}1.01 & 2.52 & 3.19 \\
& & 200  & \cellcolor{bestyellow}-0.64 & -7.21 & -2.72 & \cellcolor{secondgreen}-1.45 & 2.76 \\
& & 500  & \cellcolor{bestyellow}-0.34 & -14.12 & -6.16 & -5.47 & \cellcolor{secondgreen}4.53 \\
& & 1000 & \cellcolor{bestyellow}-0.15 & -18.33 & -8.36 & -7.78 & \cellcolor{secondgreen}6.65 \\
\midrule
\multirow{6}{*}{Scaled Beta $(5,5)$}
& \multirow{6}{*}{Symmetric, strong peak}
& 30   & \cellcolor{bestyellow}-4.82 & 74.77 & 37.17 & \cellcolor{secondgreen}36.02 & 45.04 \\
& & 50   & \cellcolor{bestyellow}-3.22 & 57.51 & 28.42 & \cellcolor{secondgreen}27.67 & 31.07 \\
& & 100  & \cellcolor{bestyellow}-1.36 & 40.36 & 19.97 & 19.89 & \cellcolor{secondgreen}19.18 \\
& & 200  & \cellcolor{bestyellow}-0.82 & 27.83 & 13.90 & 13.97 & \cellcolor{secondgreen}11.97 \\
& & 500  & \cellcolor{bestyellow}-0.20 & 15.02 & 8.09 & 8.14 & \cellcolor{secondgreen}7.25 \\
& & 1000 & \cellcolor{bestyellow}-0.22 & 7.40 & \cellcolor{secondgreen}4.47 & 4.52 & 5.55 \\
\midrule
\multirow{6}{*}{Scaled Beta $(0.8,0.2)$}
& \multirow{6}{*}{J-shaped}
& 30   & \cellcolor{secondgreen}-6.50 & \cellcolor{bestyellow}-4.09 & -38.69 & -12.90 & -11.76 \\
& & 50   & \cellcolor{bestyellow}-4.51 & \cellcolor{secondgreen}-14.33 & -40.37 & -18.25 & -18.38 \\
& & 100  & \cellcolor{bestyellow}-1.88 & -24.01 & -42.06 & -23.11 & \cellcolor{secondgreen}-22.84 \\
& & 200  & \cellcolor{bestyellow}-1.28 & -31.32 & -43.37 & -26.81 & \cellcolor{secondgreen}-24.73 \\
& & 500  & \cellcolor{bestyellow}-0.68 & -38.48 & -44.49 & -29.98 & \cellcolor{secondgreen}-24.40 \\
& & 1000 & \cellcolor{bestyellow}-0.17 & -42.69 & -45.24 & -31.71 & \cellcolor{secondgreen}-23.25 \\
\midrule
\multirow{6}{*}{Scaled Beta $(0.2,0.8)$}
& \multirow{6}{*}{Reverse J-shaped}
& 30   & \cellcolor{bestyellow}-6.42 & 29.85 & -12.45 & -12.88 & \cellcolor{secondgreen}-11.85 \\
& & 50   & \cellcolor{secondgreen}-4.47 & \cellcolor{bestyellow}2.73 & -22.33 & -19.07 & -19.07 \\
& & 100  & \cellcolor{bestyellow}-2.71 & \cellcolor{secondgreen}-17.91 & -28.69 & -23.60 & -23.35 \\
& & 200  & \cellcolor{bestyellow}-1.11 & -29.88 & -32.07 & -26.32 & \cellcolor{secondgreen}-24.14 \\
& & 500  & \cellcolor{bestyellow}-0.50 & -40.16 & -36.51 & -29.93 & \cellcolor{secondgreen}-24.30 \\
& & 1000 & \cellcolor{bestyellow}-0.59 & -45.47 & -38.74 & -31.98 & \cellcolor{secondgreen}-23.68 \\
\bottomrule
\end{tabular}
\end{table}

\begin{table}[H]
\centering
\caption{Percentage relative errors of estimating SD from quartiles with a fixed population minimum using different shapes of scaled beta distributions compared with existing three-number methods. The smallest absolute error in each row is highlighted in yellow and the second smallest is highlighted in green. Note: Scaled Beta $= L + (U-L)\times \text{Beta}(\text{shape}_1,\text{shape}_2)$ with fixed $L=18$ and $U=120$.}
\label{tab:gbeta_4p}
\setlength{\tabcolsep}{8pt}
\begin{tabular}{@{}llrrrrr@{}}
\toprule
Distribution & Shape remark & $n$ & QE-beta & QE\_3 & BC\_3 & Wan\_3 \\
\midrule

\multirow{6}{*}{Scaled Beta $(1,1)$}
& \multirow{6}{*}{Uniform}
& 30   & \cellcolor{bestyellow}4.59 & 28.03 & \cellcolor{secondgreen}8.27 & 25.65 \\
& & 50   & \cellcolor{bestyellow}5.51 & 30.56 & \cellcolor{secondgreen}8.94 & 26.77 \\
& & 100  & \cellcolor{bestyellow}5.26 & 30.78 & \cellcolor{secondgreen}10.53 & 28.12 \\
& & 200  & \cellcolor{bestyellow}3.54 & 29.13 & \cellcolor{secondgreen}9.75 & 28.20 \\
& & 500  & \cellcolor{bestyellow}1.50 & 27.34 & \cellcolor{secondgreen}9.03 & 28.12 \\
& & 1000 & \cellcolor{bestyellow}0.59 & 27.51 & \cellcolor{secondgreen}8.02 & 28.32 \\
\midrule

\multirow{6}{*}{Scaled Beta $(0.5,0.5)$}
& \multirow{6}{*}{U-shaped}
& 30   & \cellcolor{bestyellow}10.43 & 56.62 & \cellcolor{secondgreen}24.09 & 42.28 \\
& & 50   & \cellcolor{bestyellow}9.62 & 55.72 & \cellcolor{secondgreen}18.46 & 45.05 \\
& & 100  & \cellcolor{bestyellow}5.94 & 51.75 & \cellcolor{secondgreen}12.68 & 46.85 \\
& & 200  & \cellcolor{bestyellow}2.31 & 46.63 & \cellcolor{secondgreen}5.30 & 47.20 \\
& & 500  & \cellcolor{bestyellow}0.50 & 43.73 & \cellcolor{secondgreen}0.86 & 47.85 \\
& & 1000 & \cellcolor{secondgreen}0.37 & 42.78 & \cellcolor{bestyellow}0.16 & 47.99 \\
\midrule

\multirow{6}{*}{Scaled Beta $(2,8)$}
& \multirow{6}{*}{Right-skewed}
& 30   &  -4.37 & 3.47 & \cellcolor{bestyellow}2.12 & \cellcolor{secondgreen}3.17 \\
& & 50   &  -4.70 & 4.25 & \cellcolor{bestyellow}2.53 & \cellcolor{secondgreen}2.60 \\
& & 100  & \cellcolor{secondgreen}-3.21 & 5.34 & 4.25 & \cellcolor{bestyellow}1.67 \\
& & 200  & \cellcolor{secondgreen}-2.71 & 6.15 & 5.62 & \cellcolor{bestyellow}1.66 \\
& & 500  & \cellcolor{bestyellow}-0.90 & 7.66 & 7.45 & \cellcolor{secondgreen}1.57 \\
& & 1000 & \cellcolor{bestyellow}-0.79 & 8.03 & 7.52 & \cellcolor{secondgreen}1.43 \\
\midrule

\multirow{6}{*}{Scaled Beta $(8,2)$}
& \multirow{6}{*}{Left-skewed}
& 30   & \cellcolor{bestyellow}1.77 & \cellcolor{secondgreen}-2.08 & 6.01 & 2.46 \\
& & 50   & \cellcolor{secondgreen}1.88 & \cellcolor{bestyellow}-0.32 & 6.82 & 2.33 \\
& & 100  & \cellcolor{secondgreen}1.46 & \cellcolor{bestyellow}0.50 & 7.10 & 1.84 \\
& & 200  & \cellcolor{bestyellow}0.77 & \cellcolor{secondgreen}0.89 & 7.31 & 1.56 \\
& & 500  & \cellcolor{bestyellow}0.52 & \cellcolor{secondgreen}1.38 & 8.55 & 1.67 \\
& & 1000 & \cellcolor{bestyellow}0.09 & \cellcolor{secondgreen}1.33 & 9.11 & 1.47 \\
\midrule
\multirow{6}{*}{Scaled Beta $(2,2)$}
& \multirow{6}{*}{Symmetric, moderate peak}
& 30   & \cellcolor{bestyellow}0.88  & 13.08 & \cellcolor{secondgreen}3.55 & 14.75 \\
& & 50   & \cellcolor{bestyellow}1.30  & 14.49 & \cellcolor{secondgreen}4.12& 14.88 \\
& & 100  & \cellcolor{bestyellow}2.40  & 15.90 & \cellcolor{secondgreen}7.11 & 15.39 \\
& & 200  & \cellcolor{bestyellow}1.59 & 15.34 & \cellcolor{secondgreen}7.28  & 15.09 \\
& & 500  & \cellcolor{bestyellow}0.97  & 14.43 & \cellcolor{secondgreen}9.19 & 14.98 \\
& & 1000 & \cellcolor{bestyellow}0.56 & 14.06 & \cellcolor{secondgreen}9.83 & 15.09 \\
\midrule
\multirow{6}{*}{Scaled Beta $(5,5)$}
& \multirow{6}{*}{Symmetric, strong peak}
& 30   & \cellcolor{bestyellow}-2.05 & \cellcolor{secondgreen}2.13 & 2.74 & 5.92 \\
& & 50   & \cellcolor{bestyellow}-1.06 & 3.86 & \cellcolor{secondgreen}3.59  & 5.70 \\
& & 100  & \cellcolor{bestyellow}-0.37 & 5.34 & \cellcolor{secondgreen}4.68  & 6.31 \\
& & 200  & \cellcolor{bestyellow}0.10 & 5.85 & \cellcolor{secondgreen}5.73  & 6.10 \\
& & 500  & \cellcolor{bestyellow}0.35 & \cellcolor{secondgreen}6.13 & 6.68 & 6.33 \\
& & 1000 & \cellcolor{bestyellow}0.19 & \cellcolor{secondgreen}5.86 & 6.70 & 6.14 \\
\midrule
\multirow{6}{*}{Scaled Beta $(0.8,0.2)$}
& \multirow{6}{*}{J-shaped}
& 30   & \cellcolor{bestyellow}-6.64  & -20.72 & -32.49 & \cellcolor{secondgreen}-16.68 \\
& & 50   & \cellcolor{bestyellow}-4.97 & -19.81 & -31.30  & \cellcolor{secondgreen}-17.42 \\
& & 100  & \cellcolor{bestyellow}-2.09 & -19.12 & -29.95 & \cellcolor{secondgreen}-17.93  \\
& & 200  & \cellcolor{bestyellow}-1.41& -19.19 & -28.91  & \cellcolor{secondgreen}-18.60  \\
& & 500  & \cellcolor{bestyellow}-0.74 & -18.82 & -28.13 & \cellcolor{secondgreen}-18.58 \\
& & 1000 & \cellcolor{bestyellow}-0.29 & -18.42 & -27.81 & \cellcolor{secondgreen}-18.30 \\
\midrule

\multirow{6}{*}{Scaled Beta $(0.2,0.8)$}
& \multirow{6}{*}{Reverse J-shaped}
& 30   & \cellcolor{bestyellow}-14.80 & 74.36 & 25.86 & \cellcolor{secondgreen}-16.30 \\
& & 50   & \cellcolor{bestyellow}-13.60 & 62.99 & \cellcolor{secondgreen}17.65 & -19.04 \\
& & 100  & \cellcolor{bestyellow}-12.10 & 59.53 & \cellcolor{secondgreen}13.55 & -18.79 \\
& & 200  & \cellcolor{bestyellow}-10.70 & 60.29 & \cellcolor{secondgreen}13.91 & -17.75 \\
& & 500  & \cellcolor{secondgreen}-10.76 & 57.09 & \cellcolor{bestyellow}10.68 & -18.41 \\
& & 1000 & \cellcolor{secondgreen}-11.90 & 54.63 & \cellcolor{bestyellow}9.46 & -18.82 \\
\bottomrule
\end{tabular}
\end{table}

\subsection{More results of PHQ-9 simulations}\label{section:phq9_more}

In this section we present further results of the simulation study where subsamples of the PHQ-9 real data are drawn.  The results are discussed in the main text.

\begin{figure}[H]
\centering
\includegraphics[width=0.7\linewidth]{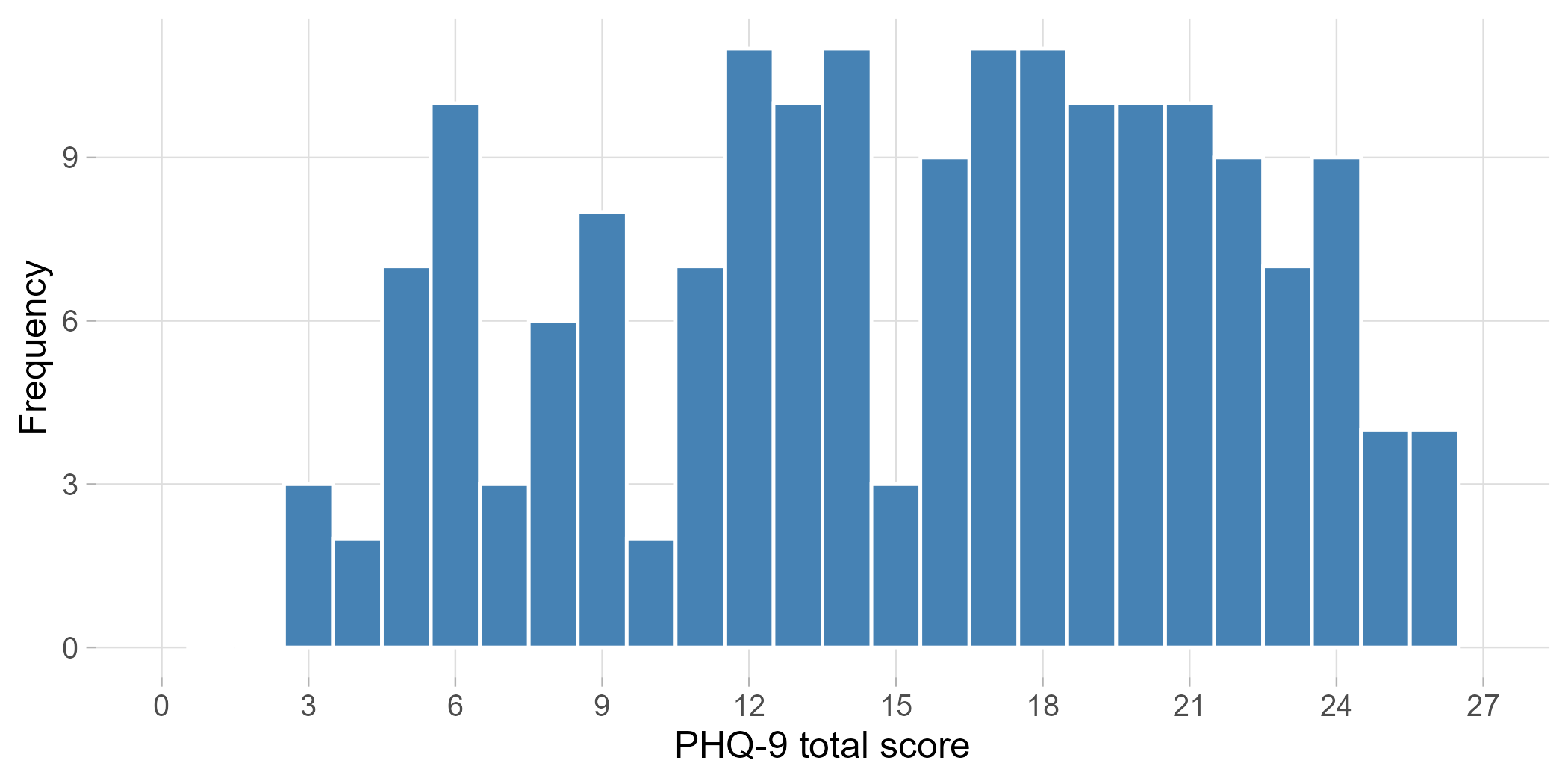}
\caption{\label{fig:phq9_hist} Histogram of PHQ-9 total score.}
\end{figure}

\begin{figure}[H]
\centering
\includegraphics[width=0.9\linewidth]{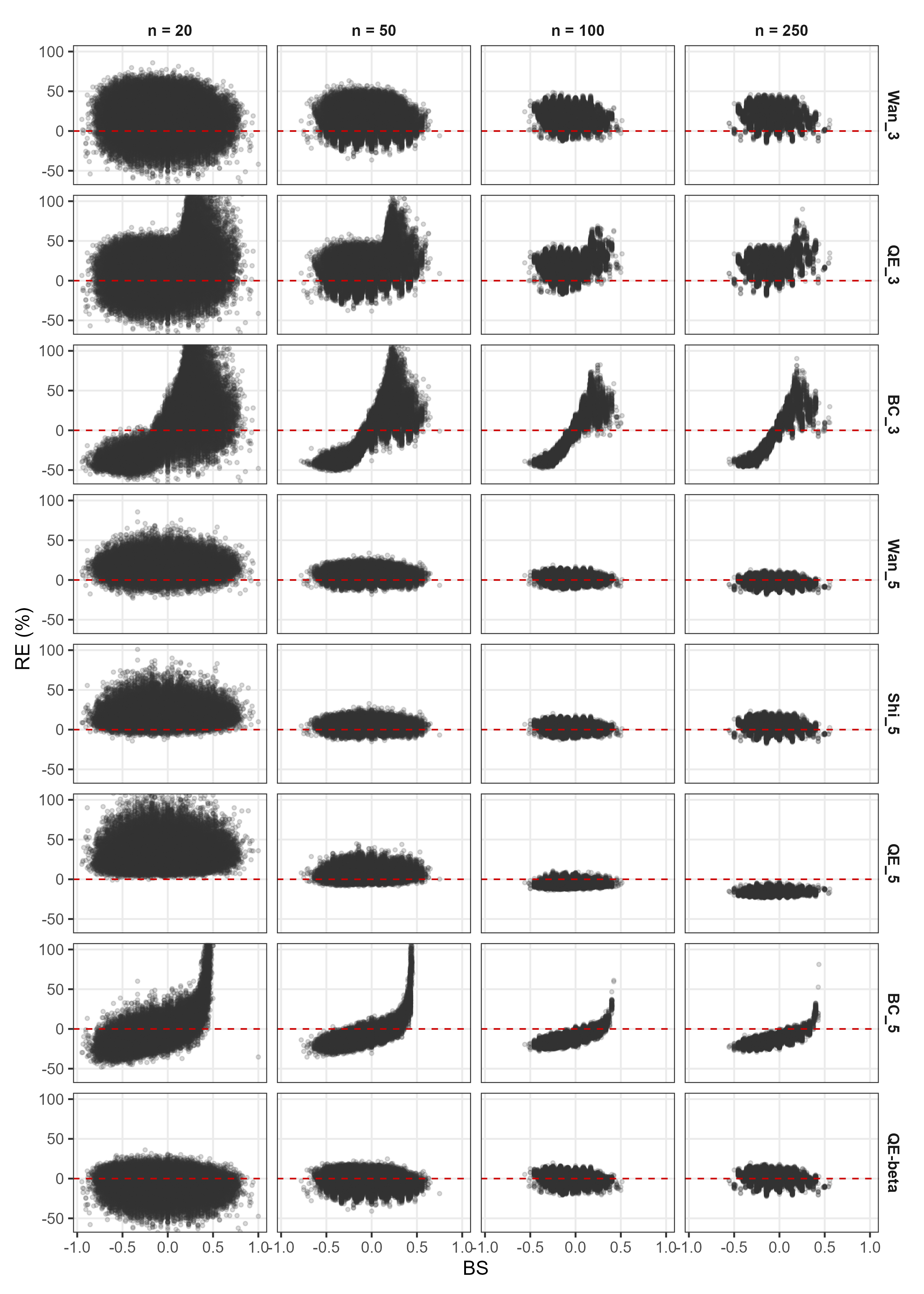}
\caption{\label{RE_vs_BS} Percentage relative errors (RE) vs Bowley skewness (BS) of estimating the standard deviation (SD) from PHQ-9 data using the proposed and existing methods for sample sizes $n=20$, $50$, $100$, and $250$.}
\end{figure}

\begin{figure}[H]
\centering
\includegraphics[width=0.85\linewidth]{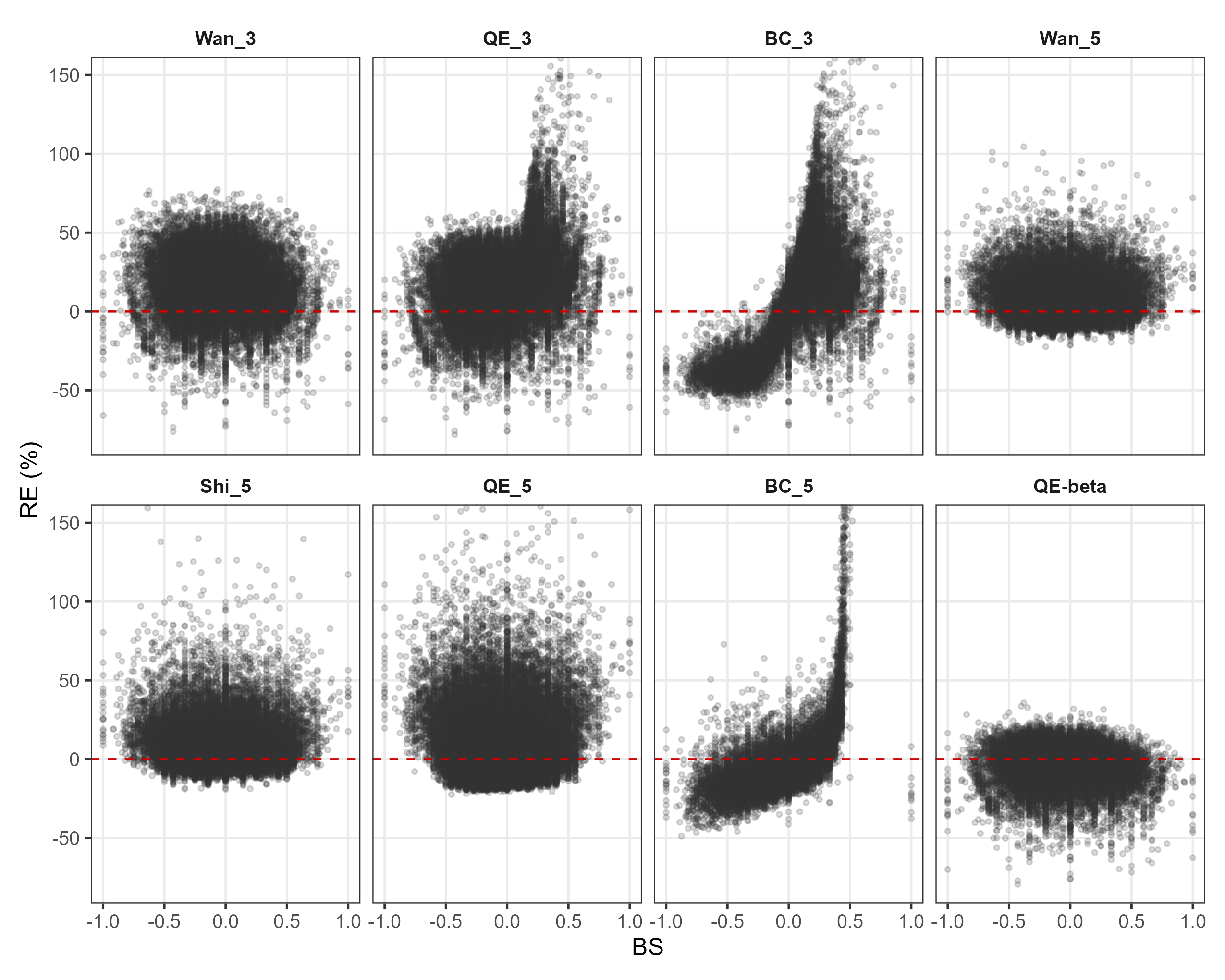}
\caption{\label{RE_vs_BS_ran_n} Percentage relative errors (RE) vs Bowley skewness (BS) of estimating the standard deviation (SD) from PHQ-9 data using the proposed and existing methods for random sample sizes $n \in [10, 200]$.}
\end{figure}

\begin{figure}[H]
\centering
\includegraphics[width=0.75\linewidth]{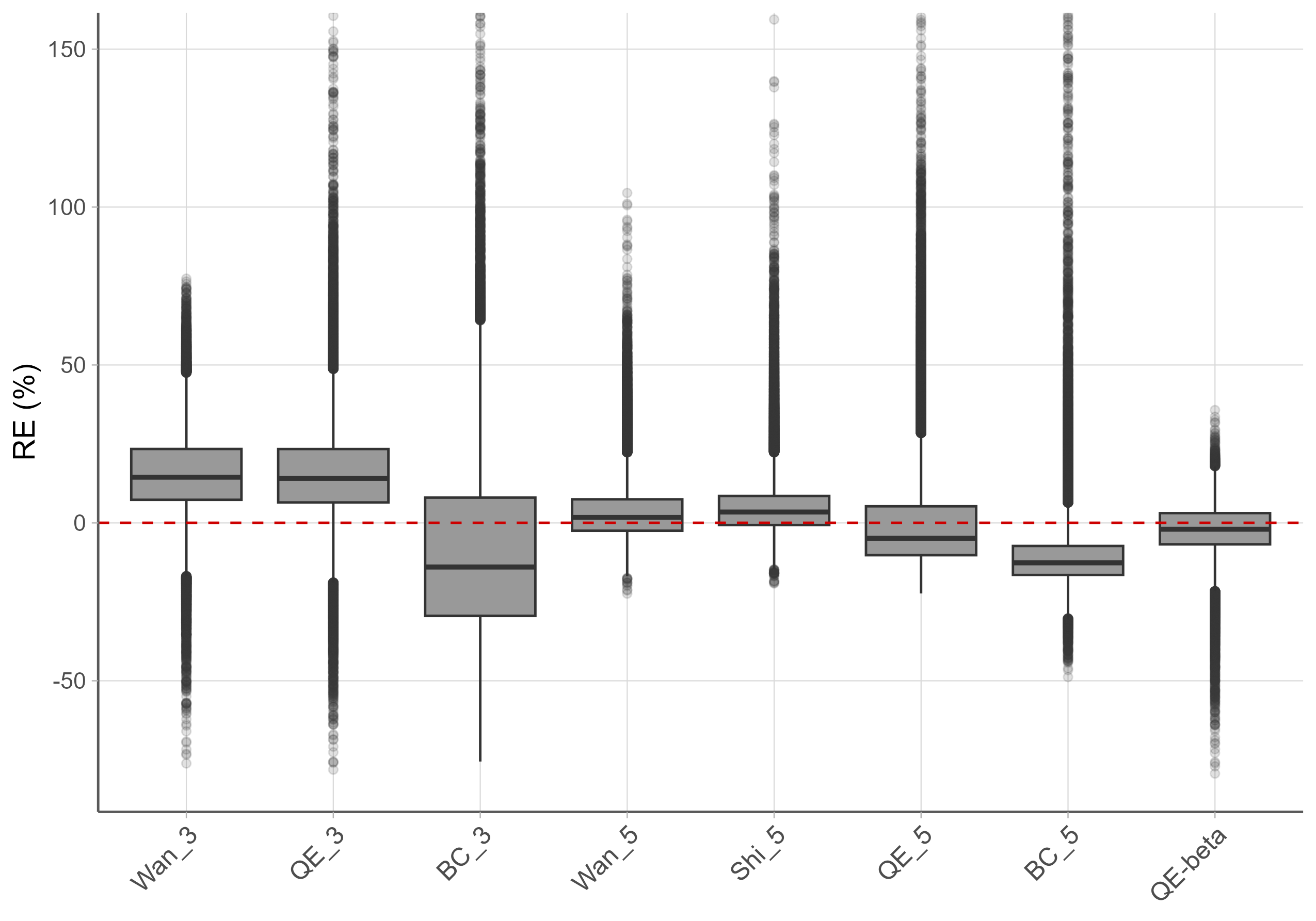}
\caption{\label{RE_ran_n} Percentage relative errors (RE) of estimating the standard deviation (SD) from PHQ-9 data using the proposed and existing methods for random sample sizes of $n \in [10, 200]$.}
\end{figure}

\begin{figure}[H]
\centering
\includegraphics[width=0.85\linewidth]{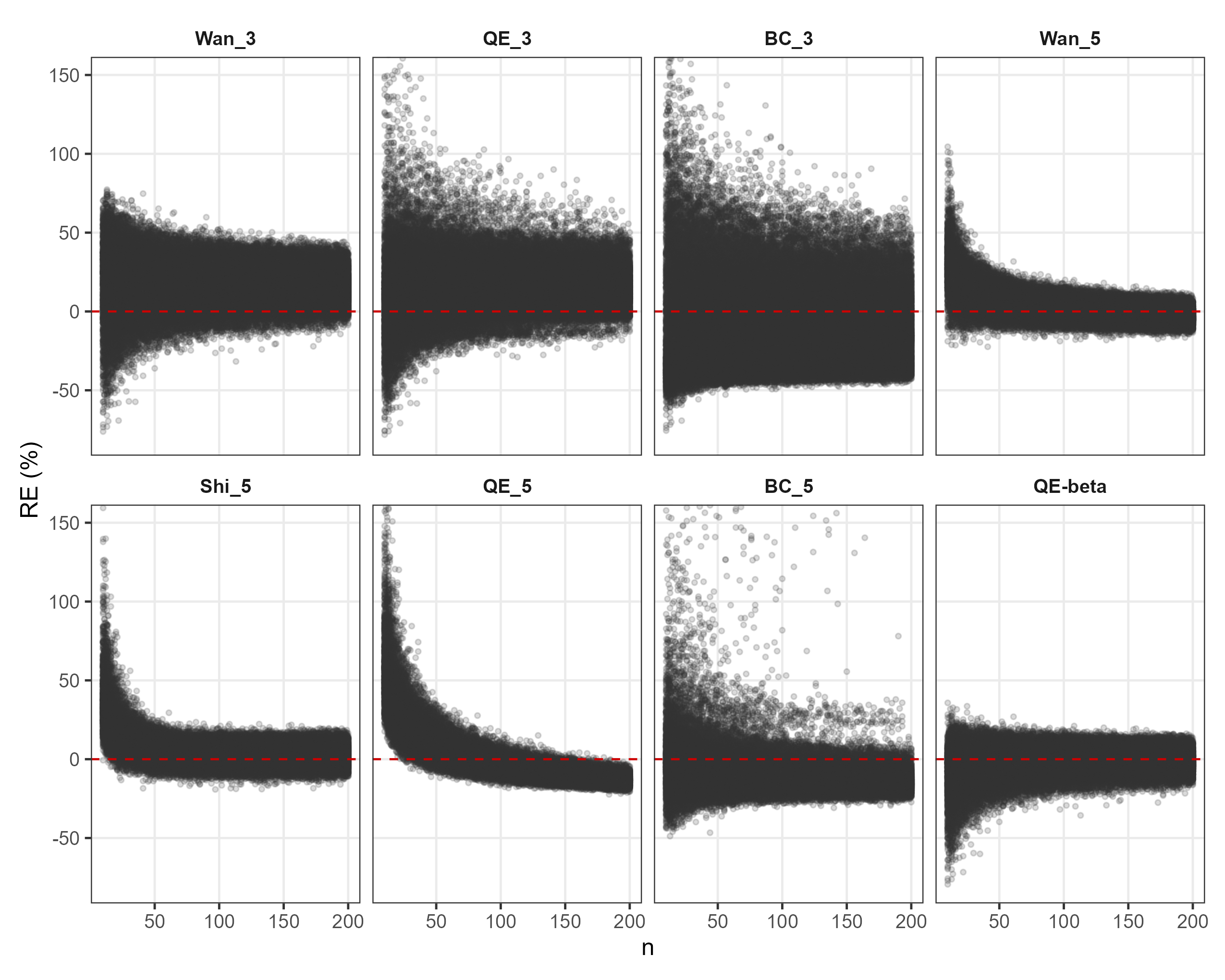}
\caption{\label{RE_vsn_ran_n} Percentage relative errors (RE) vs sample size (n) of estimating the standard deviation (SD) from PHQ-9 data using the proposed and existing methods for random sample sizes of $n \in [10, 200]$.}
\end{figure}

\end{appendix}

\end{document}